\pdfoutput=1
\documentclass[11pt,twoside,a4paper,cmspaper,final,collab]{cms-tdr}
\def\svnVersion{4695184a}\def\svnDate{2024/01/20}\def\cmsCernNoTag{CERN-EP-2023-169}\def\cmsCernDate{\today}\def\cmsMessage{Published in the Journal of High Energy Physics as \href{http://dx.doi.org/10.1007/JHEP01(2024)101}{\doi{10.1007/JHEP01(2024)101}.}}
\begin{document}\cmsNoteHeader{SMP-18-010}

\newlength\cmsFigWidth
\setlength\cmsFigWidth{0.65\textwidth}

\newlength\cmsTabSkip\setlength{\cmsTabSkip}{1ex}
\providecommand{\cmsTable}[1]{\resizebox{\textwidth}{!}{#1}}
\newcommand{\sintwoth}{\ensuremath{\sin^{2}\theta_{\mathrm{W}}^\text{eff}}\xspace}
\DeclareRobustCommand{\tauleft}{{\HepParticle{\uptau}{}{\mathrm{L}}}\xspace}
\DeclareRobustCommand{\tauright}{{\HepParticle{\uptau}{}{\mathrm{R}}}\xspace} 
\DeclareRobustCommand{\tauleftright}{{\HepParticle{\uptau}{}{\mathrm{L/R}}}\xspace}
\DeclareRobustCommand{\taurightleft}{{\HepParticle{\uptau}{}{\mathrm{R/L}}}\xspace}
\DeclareRobustCommand{\PaDo}{{\HepParticle{\Pa}{1}{}}\Xspace}
\DeclareRobustCommand{\PaDom}{{\HepParticle{\Pa}{1}{-}}\Xspace}

\newcommand{\taupi}{\ensuremath{\PGtm\to\PGpm\PGn}\xspace}
\newcommand{\tauthreepi}{\ensuremath{\PGtm\to\PGpm\PGpp\PGpm\PGn}\xspace}
\DeclareRobustCommand{\PGrm}{{\HepParticle{\uprho}{}{-}}\xspace}
\newcommand{\taurho}{\ensuremath{\PGtm\to\PGrm\PGn}\xspace}
\newcommand{\rhopipi}{\ensuremath{\PGrm\to\PGpm\PGpz}\xspace}

\DeclareRobustCommand{\tauleftminus}{{\HepParticle{\uptau}{\mathrm{L}}{-}}\xspace}
\DeclareRobustCommand{\tauleftplus}{{\HepParticle{\uptau}{\mathrm{L}}{+}}\xspace}

\DeclareRobustCommand{\taurightminus}{{\HepParticle{\uptau}{\mathrm{R}}{-}}\xspace}
\DeclareRobustCommand{\taurightplus}{{\HepParticle{\uptau}{\mathrm{R}}{+}}\xspace}

\DeclareRobustCommand{\tauleftrightminus}{{\HepParticle{\uptau}{\mathrm{L/R}}{-}}\xspace}
\DeclareRobustCommand{\taurightleftplus}{{\HepParticle{\uptau}{\mathrm{R/L}}{+}}\xspace}

\newcommand{\tautau}{\ensuremath{\PGtp\PGtm}\xspace}

\newcommand{\zll}{\ensuremath{\PZ\to\ell^+\ell^-}\xspace}
\newcommand{\zee}{\ensuremath{\PZ\to\Pep\Pem}\xspace}
\newcommand{\gammatautau}{\ensuremath{\gamma\to\tautau}\xspace}
\newcommand{\ztautau}{\ensuremath{\PZ\to\tautau}\xspace}
\newcommand{\dytautau}{\ensuremath{\PZ/\gamma\to\tautau}\xspace}
\newcommand{\ztautaunegpol}{\ensuremath{\PZ\to\tauleftminus\taurightplus}\xspace}
\newcommand{\ztautaupospol}{\ensuremath{\PZ\to\taurightminus\tauleftplus}\xspace}
\newcommand{\ztautauposnegpol}{\ensuremath{\PZ\to\tauleftrightminus\taurightleftplus}\xspace}  

\newcommand{\polarisation}{\ensuremath{\mathcal{P}_\tau}\xspace}
\newcommand{\averagepolarisation}{\ensuremath{\langle\polarisation\rangle}\xspace}

\newcommand{\taulep}{\ensuremath{\PGt_\ell}\xspace}
\newcommand{\tauel}{\ensuremath{\PGt_\Pe}\xspace}
\newcommand{\taumu}{\ensuremath{\PGt_\PGm}\xspace}
\newcommand{\tauhad}{\ensuremath{\PGt_\mathrm{h}}\xspace}
\newcommand{\muhad}{\ensuremath{\taumu\tauhad}\xspace}
\newcommand{\elhad}{\ensuremath{\tauel\tauhad}\xspace}
\newcommand{\hadhad}{\ensuremath{\tauhad\tauhad}\xspace}
\newcommand{\elmu}{\ensuremath{\tauel\taumu}\xspace}
\newcommand{\taulepton}{\PGt lepton\xspace}
\newcommand{\tauleptonminus}{\PGtm lepton\xspace}
\newcommand{\tauleptons}{\PGt leptons\xspace}
\newcommand{\tauleptonsminus}{\PGtm leptons\xspace}

\newcommand{\emCombinedOneprongOneprong}{\ensuremath{\Pe+\PGm}\xspace}

\newcommand{\etCombinedAoneOneprong}{\ensuremath{\Pe+\PaDo}\xspace}
\newcommand{\etCombinedRhoOneprong}{\ensuremath{\Pe+\PGr}\xspace}
\newcommand{\etCombinedOneprongOneprong}{\ensuremath{\Pe+\PGp}\xspace}

\newcommand{\mtCombinedAoneOneprong}{\ensuremath{\PGm+\PaDo}\xspace}
\newcommand{\mtCombinedRhoOneprong}{\ensuremath{\PGm+\rho}\xspace}
\newcommand{\mtCombinedOneprongOneprong}{\ensuremath{\PGm+\PGp}\xspace}

\newcommand{\ttRho}{\ensuremath{\rho+\tauhad}\xspace}
\newcommand{\ttCombinedAoneAone}{\ensuremath{\PaDo+\PaDo}\xspace}
\newcommand{\ttCombinedAoneOneprong}{\ensuremath{\PaDo+\PGp}\xspace}
\newcommand{\ttCombinedOneprongOneprong}{\ensuremath{\PGp+\PGp}\xspace}

\newcommand{\pol}{\ensuremath{\langle\mathcal{P}_{\PGt}\rangle}\xspace}

\cmsNoteHeader{SMP-18-010}

\title{Measurement of the \texorpdfstring{\PGt}{tau} lepton polarization in \texorpdfstring{\PZ}{Z} boson decays in proton-proton collisions at  \texorpdfstring{$\sqrt{s} = 13\TeV$}{sqrt(s) = 13 TeV}}

\date{\today}

\abstract{The polarization of \tauleptons is measured using leptonic and hadronic \taulepton decays in \ztautau events in proton-proton collisions at $\sqrt{s}=13\TeV$ recorded by CMS at the CERN LHC with an integrated luminosity of 36.3\fbinv. The measured \tauleptonminus polarization at the \PZ boson mass pole is $\mathcal{P}_{\PGt}(\PZ)=-0.144\pm0.006\stat\pm0.014\syst=-0.144\pm0.015$, in good agreement with the measurement of the \taulepton asymmetry parameter of $A_{\PGt}=0.1439\pm0.0043=-\mathcal{P}_{\PGt}(\PZ)$ at LEP. The \taulepton polarization depends on the ratio of the vector to axial-vector couplings of the \PGt leptons in the neutral current expression, and thus on the effective weak mixing angle $\sin^{2}\theta_{\PW}^{\text{eff}}$, independently of the \PZ boson production mechanism. The obtained value $\sin^{2}\theta_{\PW}^{\text{eff}}=0.2319\pm0.0008\stat\pm0.0018\syst=0.2319\pm0.0019$ is in good agreement with measurements at $\Pep\Pem$ colliders. }

\hypersetup{%
pdfauthor={CMS Collaboration},%
pdftitle={Measurement of tau lepton polarization in Z boson decays},%
pdfsubject={CMS},%
pdfkeywords={CMS, hadron-hadron scattering, tau physics}
}

\maketitle

\section{Introduction}
Measuring standard model (SM) parameters with high precision in various processes may reveal yet unknown physics phenomena as deviations from SM predictions.
One of the fundamental parameters is the effective weak mixing angle \sintwoth, which leads to 
different couplings for right- and left-handed fermions in weak neutral currents. 
A consequence of this difference is the effective
polarization of fermion-antifermion pairs in \PZ boson decays. 
The \tauleptonminus polarization is defined as $\mathcal{P}_{\PGt} = (\sigma_{+} - \sigma_{-}) / (\sigma_{+} + \sigma_{-})$, where $\sigma_{+}$ and $\sigma_{-}$
are the cross sections  for the production of $\PGtm$ leptons with positive and negative helicities, respectively. 

The helicity of \tauleptons from \PZ boson decays can be measured from energy and angular distributions of the 
\taulepton decay products. 
The polarization measures the ratio of vector to axial-vector neutral current
couplings of the \taulepton, 
and therefore this ratio provides a further
measurement of the weak mixing angle solely from \taulepton couplings as detailed in the following.

\begin{figure}[hbtp]
\centering
\includegraphics[width=0.45\textwidth]{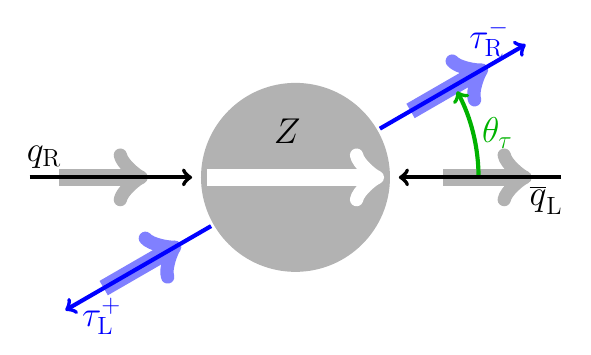}
\hfill
\includegraphics[width=0.45\textwidth]{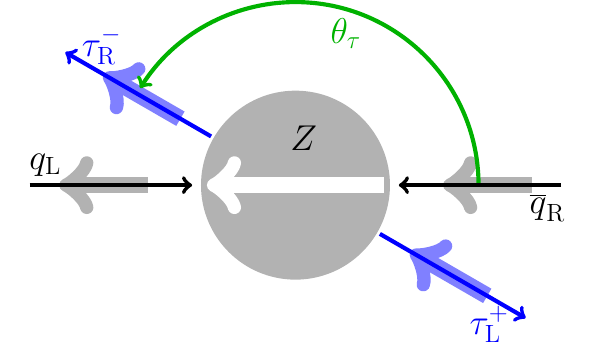}
\\
\includegraphics[width=0.45\textwidth]{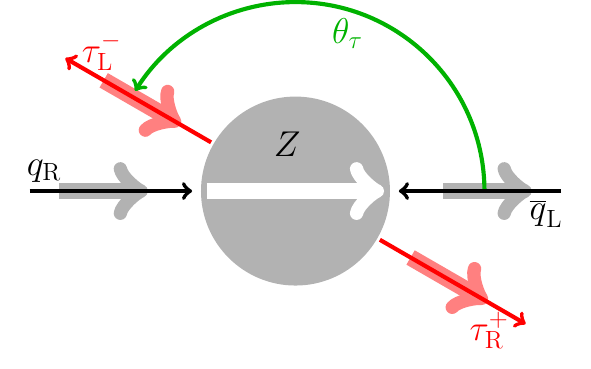}
\hfill
\includegraphics[width=0.45\textwidth]{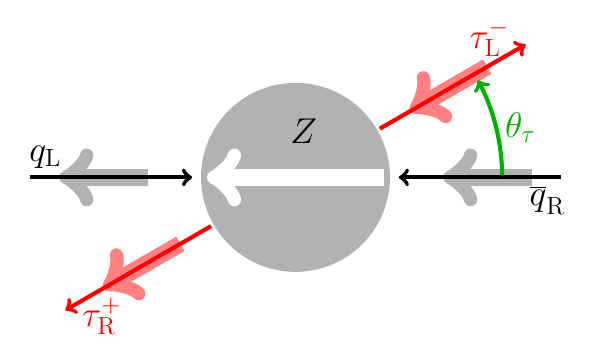}
\caption{The four possible helicity states of incoming quarks and outgoing \tauleptons. Thin arrows depict the 
direction of movement and the thick arrows show the spin of the particles. The angle $\theta_{\PGt}$ is the scattering angle of the $\PGtm$ lepton with respect to the quark momentum in the rest frame of the \PZ boson.}
\label{figure_qq_Z_tautau_helicity_states}
\end{figure}

In \ztautau events, the helicity of the $\PGtp$ lepton is expected to have a sign opposite to that of the $\PGtm$ helicity.
The differential cross section  for the process $\Pq\Paq\to\PZ\to\PGtp\PGtm$ can be expressed at the lowest order~\cite{Eberhard:201688}:
\begin{linenomath*}\begin{equation}\label{crossection}
\frac{\rd\sigma}{\rd\cos\theta_{\PGt}} = F_{0}(\hat{s})(1+\cos^{2}\theta_{\PGt}) + 2F_{1}(\hat{s})\cos\theta_{\PGt} - \lambda_{\PGt}[F_{2}(\hat{s})(1+\cos^{2}\theta_{\PGt}) + 2F_{3}(\hat{s})\cos\theta_{\PGt}].
\end{equation}\end{linenomath*}
Here $\theta_{\PGt}$ is the scattering angle of the $\PGtm$ with respect to the quark momentum in the rest frame of the 
\PZ boson, as illustrated in Fig.~\ref{figure_qq_Z_tautau_helicity_states}. The symbol $\lambda_{\PGt}$ denotes the sign of positive or negative helicity of the $\PGtm$ lepton and
$\hat{s}$ is the squared center-of-mass energy of the $\Pq\Paq$ pair. 
The $F_{i}(\hat{s})$ are structure functions for the initial and final fermion pairs describing the strength and shape of the  \PZ resonance and its dependence on the
vector and axial-vector neutral-current coupling constants.
The total cross section is: 
\begin{linenomath*}\begin{equation}\label{eq12}
\sigma  = \sum\limits_{\lambda_{\PGt}=\pm1} \int\frac{\rd\sigma}{\rd\cos\theta_{\PGt}}\rd\cos\theta_{\PGt}. 
\end{equation}\end{linenomath*}
From the cross section in Eq.~(\ref{crossection}), the following quantities can be determined: the asymmetry $A_\text{FB}$  
 of the forward-backward cross sections, the polarization $\mathcal{P}_{\PGt}$ of the $\PGtm$ lepton,
and  the forward-backward polarization asymmetry $A_\text{FB}^\text{pol}$:
\begin{linenomath*}\begin{equation} \label{Z_Assymetries}
\begin{split}
&A_\text{FB} = \frac{1}{\sigma}[\sigma(\cos\theta_{\PGt} > 0) - \sigma(\cos\theta_{\PGt} < 0)] = \frac{3F_{1}(\hat{s})}{4F_{0}(\hat{s})},\\
&\mathcal{P}_{\PGt} = \frac{1}{\sigma}[\sigma(\lambda_{\PGt} = +1) - \sigma(\lambda_{\PGt} = -1)] = -\frac{F_{2}(\hat{s})}{F_{0}(\hat{s})}, \\
&A_\text{FB}^\text{pol} = \frac{1}{\sigma}[\mathcal{P}_{\PGt}(\cos\theta_{\PGt} > 0) - \mathcal{P}_{\PGt}(\cos\theta_{\PGt} < 0)] = -\frac{3F_{3}(\hat{s})}{4F_{0}(\hat{s})}, 
\end{split}
\end{equation}\end{linenomath*}
where the symbols $F_{i}$ correspond to the structure functions in Eq.~(\ref{crossection}).

The cross section includes the contributions from
\PZ boson exchange, photon exchange, and photon-\PZ interference.
If $\sqrt{\hat{s}}$ is equal to the Z boson mass $M_{\PZ}$, the contributions from photon exchange cancel in the numerator of the asymmetries and the following holds:
\begin{linenomath*}\begin{equation} \label{consts1}
\begin{split}
&A_\text{FB} = \frac{3}{4}A_{\text{f}}A_{\PGt},\\
&\mathcal{P}_{\PGt}  =  -A_{\PGt}, \\
&A_\text{FB}^\text{pol} = \frac{3}{4}A_{\text{f}},
\end{split}
\end{equation}\end{linenomath*}
where $A_{\PGt} = 2v_{\PGt}a_{\PGt}/(v^{2}_{\PGt} + a^{2}_{\PGt})$ and $A_{\text{f}} = 2v_{\text{f}}a_{\text{f}}/(v^{2}_{\text{f}} + a^{2}_{\text{f}})$ are the asymmetry parameters
defined by the effective neutral current vector and axial-vector couplings
$v_{\PGt}$ and $a_{\PGt}$ for the \taulepton, and $v_{\text{f}}$ and $a_{\text{f}}$, the parameters for the initial-state fermions, respectively. 
In the limit $v_{\PGt} \ll a_{\PGt}$, the polarization can be written as $\mathcal{P}_{\PGt} \approx -2v_{\PGt}/a_{\PGt}$, and is simply related to \sintwoth:
\begin{linenomath*}\begin{equation}\label{eqnumber5}
\mathcal{P}_{\PGt}=-A_{\PGt} = -\frac{2v_{\PGt}a_{\PGt}}{v^{2}_{\PGt} + a^{2}_{\PGt}} \approx -2 \frac{v_{\PGt}}{a_{\PGt}} = -2(1 - 4\sintwoth).
\end{equation}\end{linenomath*}
Hence, a measurement of the polarization $\mathcal{P}_{\PGt}$ can provide a precise determination of \sintwoth
using only \taulepton couplings.
Comparison with the value of \sintwoth measured in the process $\Pep\Pem\to\PZ$ at LEP~\cite{Heister:2001uh,Abreu:1999wv,Acciarri:1998vg,Abbiendi:2001km}
tests the lepton universality of the weak neutral current.

Since \tauleptons decay rapidly inside the detector, their polarization is measured by analyzing the energy and direction of their decay products. 
Observables used for this analysis are the angles and momenta of the daughter particles with respect to the 
boost direction of the \taulepton or intermediate particles in the decay of the \taulepton.

Measurements of the \taulepton polarization have been performed in $\Pep\Pem$ annihilation by the four LEP experiments at center-of-mass energies
near $M_{\PZ}$~\cite{Heister:2001uh,Abreu:1999wv,Acciarri:1998vg,Abbiendi:2001km}, at the linear collider experiment SLD~\cite{PhysRevLett.86.1162}, and
in proton-proton ($\Pp\Pp$) collisions by the ATLAS experiment at the CERN LHC at $\sqrt{s}=8\TeV$~\cite{Aaboud:2017lhv}.

The measurement of the \taulepton polarization in \PZ boson decays was pioneered at LEP~\cite{Eberhard:201688}, where the polarization could be measured in a very small window around the \PZ pole and as a function of the polar emission angle $\cos\theta_{\PGt}$ of the \taulepton with respect to the incoming electrons. A maximum likelihood fit to this angular dependence provides the best measurement of the polarization. 

A complication occurs when the polarization is measured in $\Pp\Pp$
collisions; in contrast to $\Pep\Pem$ collisions, the polar emission angle $\theta_{\PGt}$ is not, or only very poorly, known and can not be used in the analysis.  
Furthermore, instead of a fixed value, measurements are always an average over a limited range of the center-of-mass energy of the $\PQd\PAQd$ and $\PQu\PAQu$ quark-antiquark pairs. The measurement is effectively an integration over the experimental width of the \PZ boson and a fairly wide range of its rapidity and transverse momentum $\pt$. The ranges are determined by the parton distribution functions (PDFs) of the proton and the acceptance of the experiment. 

In this paper, a measurement of the \tauleptonminus polarization $\mathcal{P}_{\PGt}$ 
is presented, based on $\PZ\to\PGt\PGt$ events in pp collisions with parton-parton center-of-mass energies
$\sqrt{\hat{s}}$ of 75--120\GeV. 
The measurement is performed by
using the following combinations of leptonic and hadronic decay modes of the \tauleptons: $\tauel\taumu$, $\tauel\tauhad$, $\taumu\tauhad$, and $\tauhad\tauhad$, where the symbol $\tauel$ refers to the decay $\PGt^-\to\Pe^-\PAGne\PGnGt$, the symbol \taumu refers to the decay $\PGt^-\to\PGm^-\PAGnGm\PGnGt$, and the $\tauhad$ to the hadronic \PGt  decay modes.

Following a short description of the CMS experiment in Section~\ref{sec_CMS} and of the Data and Monte Carlo samples (Section~\ref{section_samples}), the 
event reconstruction and the event selection are detailed in Sections~\ref{sec_evt_reco} and~\ref{sec_evt_sel}, respectively. 
The background estimates are described in Section~\ref{sec_bkg}, before the \taulepton spin variables are discussed in Section~\ref{section_optimal_observables}. 
After a discussion of systematic uncertainties in Section~\ref{sec_syst}, the measurement of the \tauleptonminus polarization $\mathcal{P}_{\PGt}$ is described in Section~\ref{section_polarisation_measurement}. 
Finally, from the measurement of $\mathcal{P}_{\PGt}$, averaged over the mass range of the \PZ boson, the effective weak mixing angle \sintwoth is derived and presented in Section~\ref{results}.  
The paper ends with a summary in Section~\ref{sec_sum}.

\section{The CMS detector}
\label{sec_CMS}
The central feature of the CMS apparatus is a superconducting solenoid of 6\unit{m} internal diameter, providing a magnetic field of 3.8\unit{T}. Within the 
magnetic volume are a silicon pixel and strip tracker, a lead tungstate crystal electromagnetic calorimeter (ECAL), 
and a brass and scintillator hadron calorimeter (HCAL), each composed of a barrel and two endcap sections. 
Forward calorimeters extend the pseudorapidity ($\eta$) coverage provided by the barrel and endcap detectors. 
Muons are detected in gas-ionization chambers embedded in the steel flux-return yoke outside the solenoid. A more detailed description of the CMS detector, together with a definition of the coordinate system used and the relevant kinematical variables, 
are reported in Ref.~\cite{CMS:2008xjf}.

Events of interest are selected using a two-tiered trigger system. The first level (L1), composed of custom hardware processors, uses information from the calorimeters and muon detectors to select events at a rate of around 100\unit{kHz} within a fixed latency of 4\mus~\cite{CMS:2020cmk}. The second level, known as the high-level trigger (HLT), consists of a farm of processors running a version of the full event reconstruction software optimized for fast processing, and reduces the event rate to around 1\unit{kHz} before data storage~\cite{CMS:2016ngn}.

\section{Data and Monte Carlo samples} \label{section_samples}
This analysis is based on $\Pp\Pp$ collision data at a center-of-mass energy of 13\TeV collected in the year 2016 with the CMS detector at the LHC.  
The analyzed data correspond to an integrated luminosity of 36.3\fbinv~\cite{CMS:2021xjt}.     

{\tolerance=1600
A Drell--Yan(DY) signal Monte Carlo (MC) sample of  $\PZ/\gamma^*\to\PGt\PGt$ events is generated for this analysis at next-to-leading order (NLO) in quantum chromodynamics (QCD), but at leading order (LO) for electromagnetic processes using \MGvATNLO v2.4.2~\cite{Alwall:2014hca}, with the hadronization step simulated by  \PYTHIA8~\cite{Sjostrand:2014zea}. 
The \textsc{CUETP8M1}~\cite{Khachatryan:2015pea} \PYTHIA tune and \PYTHIA8 v8.226 are used.
The \PGt polarization flag \textsc{TauDecays:externalMode = 0} is used to ensure proper \PGt decay simulation by \TAUOLA 1.1.6~\cite{Jadach:1993hs, Davidson:2010rw} following the helicity assignment of \MGvATNLO.
The NNPDF3.0~\cite{NNPDF:2014otw} PDFs are used. A sample of about 60 million events was produced in this way and processed through the detector simulation and reconstruction chain. The simulation of the detector response is based on \GEANTfour~\cite{GEANT4:2002zbu}.

Differences in the mass and $\pt$ distributions of the \PGt pair between data and simulations are observed  and 2D weights based on these variables are derived and applied to simulated Drell--Yan events~\cite{2018283}. 
Since \MGvATNLO is NLO in QCD, additional jets are included at the matrix element level. 
A reweighting between 3.9 and 12.3\% was applied to the distribution of the number of generated partons in the event to match the distribution of jets. The analysis is not sensitive to those additional jets.

The main background processes, such as $\PW$+jets and other Drell--Yan processes like $\PZ/\gamma^*\to\Pe\Pe$ and $\PZ/\PGg^*\to\PGm\PGm$, are generated at LO for electromagnetic processes with \MGvATNLO and interfaced with \PYTHIA8 with the tune \textsc{CUETP8M1}~\cite{Khachatryan:2015pea}.  
Backgrounds from \textit{s}-channel and $\PQt\PW$ single top quark associated production are generated 
with \POWHEG~v1.0~\cite{Re:2010bp} and \POWHEG~v2.0~\cite{Alioli:2009je}, respectively, interfaced to \PYTHIA8 with the \textsc{CUETP8M1} tune. The top quark pair production ($\cPqt\cPaqt$) sample is produced with \POWHEG~v1.0~\cite{Frixione:2007nw} interfaced 
to \PYTHIA8 with the \textsc{CUETP8M2T4}~\cite{CMS:2016kle} tune. 
Diboson samples are generated with \MGvATNLO and tune \textsc{CUETP8M1},  interfaced to \PYTHIA8 and \PGt lepton decays are processed by \TAUOLA. Again, the NNPDF3.0~\cite{NNPDF:2014otw} PDFs are used for the background processes.
\par}

The impact of multiple $\Pp\Pp$ collisions in the same or adjacent bunch crossings (pileup) on event reconstruction~\cite{CMS:2020ebo} is accounted for in simulation by superimposing simulated minimum bias pp events on top of each process of interest.  
Because the distribution of the number of pileup events in the simulation is not the same as in data, the simulation is weighted to match the data.

\section{Event reconstruction \label{sec_evt_reco}}

The particle-flow (PF) algorithm~\cite{Sirunyan:2017ulk} reconstructs and identifies each individual particle in an event, with an optimized combination of information from the various elements of the CMS detector. The energy of photons is obtained from the ECAL measurement. The energy of electrons is determined from a combination of the electron momentum at the primary vertex (PV) as determined by the tracker, the energy of a corresponding ECAL cluster, and the energy sum of all bremsstrahlung photons spatially compatible with originating from the electron track. The energy of muons is obtained from the curvature of the corresponding track. The energy of charged hadrons is determined from a combination of their momentum measured in the tracker and the matching ECAL and HCAL energy deposits, corrected for the response function of the calorimeters to hadronic showers. 
Finally, the energy of neutral hadrons is obtained correspondingly from the corrected ECAL and HCAL energies.

The PV is taken to be the vertex corresponding to the hardest scattering in the event, evaluated from the largest value of summed physics-object $\pt^2$ as described in Ref.~\cite{2018283}.

For each event, hadronic jets are clustered from these reconstructed particles based on the infrared- and collinear-safe anti-\kt algorithm~\cite{Cacciari:2008gpt,Cacciari:2011ma} with a distance parameter of 0.4. The jet momentum is determined as the vector sum of all particle momenta in the jet, and is found from simulation to be, on average, within 5-10\% of the generated momentum over the entire \pt spectrum and detector acceptance. 
Pileup interactions can lead to additional tracks and calorimetric energy depositions to the jet momentum. To mitigate this effect, charged particles identified to be originating from pileup vertices are discarded, and an offset correction is applied to correct for remaining contributions. Jet energy corrections are derived from simulation to bring the measured response of jets to that of particle level jets on average. In situ measurements of the momentum balance in dijet, photon+jet, $\PZ$+jet, and multijet events are used to account for any residual differences in the jet energy scale between data and simulation. The jet energy resolution amounts typically to 15--20\% at 30\GeV, 10\% at 100\GeV, and 5\% at 1\TeV~\cite{CMS:2016lmd}. Additional selection criteria are applied to each jet to remove jets potentially dominated by anomalous contributions from various subdetector components or reconstruction failures.

Electrons are reconstructed within the geometric acceptance $\abs{\eta} < 2.5$. The momentum resolution for electrons with $\pt \approx 45\GeV$ from \zee decays ranges from 1.6 to 5.0\%. It is generally better in the barrel region than in the endcaps, and also depends on the bremsstrahlung energy emitted by the electron before reaching the ECAL~\cite{CMS:2020uim}.

Muons are measured in the range $\abs{\eta}<2.4$, with detection planes made out of three types of gas-ionization detectors: drift tubes, cathode strip chambers, and 
resistive-plate chambers.  
Matching muons to tracks measured in the silicon tracker results in a relative transverse momentum resolution of 1\% in the barrel and 3\% in the endcaps for muons with \pt up to 100\GeV~\cite{CMS:2012nsv}.

The missing transverse momentum vector \ptvecmiss in the event is defined as the negative vector 
sum of the momenta of all reconstructed particles in an event projected onto the plane perpendicular to the beam axis. 
It is computed from the PF candidates weighted by their probability to originate from the PV~\cite{CMS:2019ctu}.
Recoil corrections are applied to account for mismodeling of \ptvecmiss in the simulated samples of
Drell--Yan and $\PW$+jets production.
The vector \ptvecmiss 
is further modified to account for corrections to the energy scale of the reconstructed jets in the event. The pileup-per-particle identification 
algorithm~\cite{Bertolini:2014bba} is applied to reduce the pileup dependence of the \ptvecmiss observable. 
The magnitude of \ptvecmiss is referred to as \ptmiss.

Leptonic \PGt decays, \tauel and \taumu, are identified as isolated electrons and muons.  
The lepton $\ell$ ($\Pe$, \PGm) isolation $I_\text{rel}(\ell)$ is defined by the following equation:
\begin{linenomath*}\begin{equation}\label{iso_lepton}
I_\text{rel}(\ell) = \frac{\sum_{\text{ch. had}} \pt + \max \left(0,\;\sum_{\text{n. had.}} \ET + \sum_{\gamma} \ET - \beta(\text{PU}) \right)}{\pt^\ell}.
\end{equation}\end{linenomath*}
In this expression, $\sum_{\text{ch. had}} \pt$
is the scalar transverse momentum sum of the charged hadrons originating from the PV within a cone of size 
$\Delta R=\sqrt{\smash[b]{(\Delta\eta)^2+(\Delta\phi)^2}}= 0.3$ or 0.4 for electrons and muons, respectively, centered on the lepton.
The quantities $\Delta\eta$ and $\Delta\phi$ measure the particle separation in $\eta$ and of the azimutal angle $\phi$ in radians.
The  sum $\sum_{\text{n. had.}} \ET + \sum_{\gamma} \ET$ in Eq.~(\ref{iso_lepton})
represents  a similar quantity for neutral particles. 

To estimate the contribution of
neutral particles due to pileup the so-called effective-area method is used for identified electrons. 
The pileup in this method is estimated as $\beta(\text{PU})$ = $\rho A_\text{eff}$, where $\rho$ is the event-specific average pileup energy
density per unit area in the $\eta-\phi$ plane and $A_\text{eff}$ is the effective area specific to the given type
of isolation.

A different method is used for the selected muons; 
the contribution of photons and neutral hadrons originating
from pileup vertices $\beta(\text{PU})$ is estimated from the scalar transverse momentum sum of charged
hadrons  originating from pileup vertices: $\beta(\text{PU})= 0.5\sum_{\text{PU}}\pt$.
The sum $\sum_\text{PU}\pt$ is multiplied
by a factor of 0.5, which corresponds approximately to the ratio of neutral-to-charged hadron
production, as estimated from simulation. 

Lepton isolation and identification efficiencies are measured with a tag-and-probe method~\cite{PhysRevLett.112.191802,CMS:2018jrd} using  \zll events in lepton \pt and $\eta$ bins from samples collected based on single-lepton triggers.

The reconstruction of the hadronically decaying \tauleptons (\tauhad candidates) is performed by the hadron-plus-strips algorithm (HPS)~\cite{CMS:2018jrd}.
Jets are reconstructed from PF constituents with the anti-\kt algorithm~\cite{Cacciari:2008gpt}
with a distance parameter of $\Delta R = 0.4$.  
They are considered as \tauhad candidates if they comprise one or three charged hadrons with a net
charge of $\pm$1, and up to two neutral pions. Dedicated attention is given to photons originating from $\PGpz\to\gamma\gamma$ 
decays likely to convert to $\Pep\Pem$ pairs, by collecting the photon and electron constituents, called strips, in an area  
around the jet direction in the $\eta-\phi$ plane. 
The size of the strip varies as a function of the \pt of the \tauhad. 
To further characterize the \tauhad signature, its visible reconstructed mass is required to be compatible with that of the 
$\PGrP{770}$ resonance for $\PGtm\to\mathrm{h}^{-}\PGpz$ signatures, and the \PaDoP{1260} resonance for  $\PGtm\to\mathrm{h}^{-}\PGpz\PGpz$ and  $\PGtm\to\mathrm{h}^{-}\mathrm{h}^{-}\mathrm{h}^{+}$  signatures, with corresponding channels for the $\PGtp$.

Energy scale corrections are computed for each decay mode using either the visible mass of both \tauleptons or the 
invariant mass distributions of the $\PGppm \PGpz$, $\PGppm \PGpz \PGpz$, and $\PGpp \PGpm \PGppm$ systems. The correction 
factors are chosen such that either the visible mass distributions of both \tauleptons
of the MC events match the data distribution, or the invariant mass distributions of the 
$\PGppm \PGpz$, $\PGppm \PGpz \PGpz$, and $\PGpp \PGpm \PGppm$ systems
match the shapes of the \PGr and $\PaDo$ resonances~\cite{CMS:2018jrd}.

In this analysis, since the decay channels of the \tauhad have 
different polarization sensitivity, it is fundamental to identify these channels correctly. 
To increase the purity of the reconstructed decay channels, a multi-class multivariate analysis (MVA)
with a boosted decision tree  algorithm (implemented using the \textsc{XGBoost} library) is applied
in addition to the HPS algorithm. 
The multivariate algorithm was developed for the analysis of the charge-conjugation and parity ($CP$) properties of the Higgs boson in \tautau decays~\cite{CMS:2021sdq}, 
which is also very sensitive to migrations between reconstructed decay modes. Based on the simulated DY signal sample, the use of the multivariate algorithm in addition to the HPS algorithm increased the purity of the $\taupi$ and $\taurho$ channels in the analysis presented here from 63 to 83\% and from 63 to 77\%, respectively. For reconstructed \tauleptons the efficiencies for the correct decay mode identifications are 76, 82 and 94\% for the decays \taupi, \taurho and \tauthreepi, respectively.

The \textsc{DeepTau} discriminator~\cite{CMS:2022prd} is used in this analysis in order to distinguish \tauh from quark and gluon jets, electrons and muons. It is a deep neural network algorithm that returns three discriminants, trained to reduce misreconstruction of jets, electrons, or muons as \tauh candidates. 
The so-called Medium Working Point (Med \textsc{DeepTau} WP) is used for  \tauh as referred to in Table~\ref{tab_sel_cuts} of the next section.

Correction factors were measured in Ref.~\cite{CMS:2018jrd} for the different working points of anti-jet, anti-electron, and 
anti-muon discriminators used to identify the \tauhad. These were measured in $\PZ\to\taumu\tauhad$ events, 
considering the visible mass of the muon or \PGt lepton candidate as observables (or alternatively the number of tracks in the signal
and isolation cones). 

\section{Event selection and categorization} \label{sec_evt_sel}

The  following analysis uses selection criteria developed for the analyses of the Higgs boson in the \tautau decay channel~\cite{2018283} 
and its $CP$ properties~\cite{CMS:2021sdq}; some of them are summarized in Table~\ref{tab_sel_cuts}.

\begin{table}[h]
  \centering
 \topcaption{Selections applied in this analysis. For the \muhad channel two triggers were used with different muon thresholds, the \pt selection threshold for \tauhad refers to both. For the \hadhad channel, the \pt selection threshold for the nonleading \tauhad was lower by 5\GeV. The label Med DeepTau WP in the last column refers to the medium working point of the DeepTau discriminator against fake \tauhad.}
  \cmsTable{
  \begin{tabular}{ c c c c c }
    \hline
 Channel & Trigger & Lepton selection & \multicolumn{2}{c}{Additional selection} \\
	\hline
        		&                          &                                                         &                          &\\
	\elmu    & $\pt^{\PGm}>8\GeV, \pt^\Pe>23\GeV$              & $\pt^\Pe>15\GeV$ , $\abs{\eta^\Pe}<2.4$                      & $I_\text{rel}(\Pe)<0.15$     &\\
	         & or $\pt^{\PGm}>23\GeV, \pt^\Pe>12\GeV$          & $\pt^{\PGm}>15\GeV$ , $\abs{\eta^{\PGm}}<2.4$                  & $I_\text{rel}(\PGm)<0.20$   &\\
		 &                          & $\pt^\ell>24\GeV$ for lead trigger leg              &                          &\\
 	         &                          &                                                         &                          &\\
	\elhad   &$ \pt^\Pe>25\GeV$                  & $\pt^\Pe>30\GeV$  , $\abs{\eta^\Pe}<2.1$                      &$I_{\mathrm{rel}}(\Pe)<0.15$   & $m_\mathrm{T}^\Pe<50\GeV$  \\
		 &                          & $\pt^{\tauhad}>30\GeV$  , $\abs{\eta^{\tauhad}}<2.3$      & Med \textsc{DeepTau} WP       &\\
	         &                          &                                                         &                          &\\
	\muhad   & $\pt^{\PGm}>22\GeV$                  & $\pt^{\PGm}>23\GeV$, $\abs{\eta^{\PGm}}<2.1, $                  & $I_{\mathrm{rel}}(\PGm)<0.15$   & $m_\mathrm{T}^{\PGm}<50\GeV$  \\
	         & or $\pt^{\PGm}>19\GeV, \pt^{\tauhad}>20\GeV$    & $\pt^{\PGm}>20\GeV$     				& Med \textsc{DeepTau} WP       &\\
	         &                          					&      $\pt^{\tauhad}>30\GeV$,  $\abs{\eta^{\tauhad}}<2.3$                                                   &                          &\\
	         &                          &                                                         &                          &\\
	\hadhad  & $\pt^{\tauhad}>35\GeV, \pt^{\tauhad}>35\GeV$  & $\pt^{\tauhad}>45(40)\GeV$ , $\abs{\eta^{\tauhad}}<2.1$       & Med \textsc{DeepTau} WP       & \\	
    \hline
 \end{tabular}}\label{tab_sel_cuts}
\end{table}

A combination of several triggers is used to cover the different channels of the \ztautau decays and described in more detail in Ref.~\cite{2018283}. The off-line selection thresholds in Table~\ref{tab_sel_cuts} are often significantly higher than the trigger threshold to ensure exactly known efficiencies. 
The pseudorapidity limits come from trigger and object reconstruction constraints. 

The electrons and muons in the \elmu channel, electron, muon and \tauhad in the \elhad, \muhad channel, respectively, and the two \tauhad in the \hadhad channel are required to be oppositely charged.
The \tauel, \taumu and \tauhad candidates are required to 
have a distance of closest approach to the PV of $\abs{d_z}<0.2\unit{cm}$ in the direction along the beam axis and
 $\abs{d_{xy}}<0.045\unit{cm}$ in the transverse plane.
 
In the \elhad channel, the trigger system requires at least one isolated electron object, whereas in the \elmu channel, the triggers rely on the presence of both an electron and a muon, allowing lower online \pt thresholds.
In the \muhad channel, events are selected with 
at least one isolated muon trigger candidate, or at least one isolated
muon and one \tauhad trigger candidate, depending on the offline muon \pt.

For the \hadhad channel, the trigger selects events with two \tauhad objects using a less restrictive isolation criteria~\cite{CMS:2018jrd} than the more selective Med \textsc{DeepTau} WP used for the analysis listed in Table~\ref{tab_sel_cuts}.

The $\PGt\PGt$ final states are categorized according to the number of identified electrons and muons
in the event. 
Events containing an electron and a muon are assigned to the \elmu category. 
If at least one \tauel and \tauhad candidate is found, but no muon, the event is assigned the \elhad final state.  Similarly the event 
is assigned to the \muhad final state if it contains at least one muon candidate, but no electron. 
If neither an electron nor a muon is found but at least two \tauhad candidates, the event is assigned to the \hadhad final state. 
Events containing three or more electrons or muons are vetoed to suppress the contribution from other Drell--Yan events. 
This ensures that the four final states are mutually exclusive and of very good purity. 

The \tauel, \taumu, and \tauhad candidates corresponding to the decay of the \taulepton pair are required to be separated by  $\Delta R>0.3$ for \elmu events and $\Delta R>0.5$ for \elhad, \muhad, \hadhad events. 
Since \tauhad candidates are wider than \taulep the required separation is slightly larger. 
The reconstructed leptons must correspond to the HLT candidates on which the trigger decision is made.
The correspondence is ensured by requiring the selected electron or muon in the \elhad and 
\muhad final states to be within a distance $\Delta R=0.5$ from the corresponding HLT objects. 
The same spatial separation is applied for both \tauhad candidates in the \hadhad final state. 

The transverse mass $\mT^\ell$ 
is defined from the transverse momentum $\pt^{\ell}$ of the electron or muon and the missing transverse momentum:
\begin{linenomath*}\begin{equation}\label{missingmass}
\mT^\ell=\sqrt{2\pt^{\ell}\pt^\text{miss}(1-\cos[\Delta\phi(\ell,\ptvecmiss)])}.
\end{equation}\end{linenomath*} 
Events containing a \PW boson are expected to have large values of \mT due to the missing neutrino, hence a selection on \mT less than 50\GeV is applied to reduce this background.

After the event selection described above, events are categorized according to the 
reconstructed decay mode 
of the hadronically decaying \tauleptons.  

\section{Background estimation} \label{sec_bkg}
The largest background comes from multijet events where one of the jets is misidentified as a \tauhad. 
Based on simulation it represents
about 84\% of the expected background in the \hadhad channel, and 16 (20\%) in the 
\elhad (\muhad) channels. 
Drell--Yan events in dilepton final states may contribute if either a lepton is identified as a \tauhad
or an additional jet is misidentified as \tauhad. 
These contributions are suppressed by rejecting events containing lepton pairs with same flavor and
opposite electric charge and applying the \textsc{DeepTau} discriminants to suppress electrons and muons wrongly reconstructed as \tauhad candidates.
Whereas control samples are used in data to evaluate the multijet background, other contributions such as \ttbar and electroweak processes (EWK: single top quark, \PZ+jets, and dibosons) rely on simulation. The $\PW$+jets background is significant in the \muhad and \elhad final states (respectively about 31 and 27\% of the expected background) and is evaluated from data as described below.   

\subsection{The \texorpdfstring{\elhad}{electron+tau} and  \texorpdfstring{\muhad}{muon+tau} final states}

The QCD multijet and $\PW$+jets background estimates are obtained by applying the so-called $ABCD$ method to four regions, delimited by the lepton ($\Pe$ or \PGm) isolation and the lepton-\tauhad pair charge combinations, opposite-sign (OS) or same-sign (SS) pairs. 
The signal region $A$ is defined by 
$I_\text{rel}(\ell)<0.15$ and an OS pair; $B$ by $I_\text{rel}(\ell)<0.15$ and an SS pair; $C$ by $I_\text{rel}(\ell)>0.15$ and an OS pair; $D$ by $I_\text{rel}(\ell)>0.15$ and an SS pair. These regions receive significant contributions from $\PW$+jets events 
that must be evaluated simultaneously with the multijet background. 
Therefore, two samples are defined in which the $ABCD$ method is used: 
one (\text{low-$\mT$}) follows the signal selection criteria for
which the transverse mass satisfies $\mT<50\GeV$; and the other (\textrm{high-$\mT$}) is enriched in $\PW$+jets events by requiring
$\mT>70\GeV$. 
The \textrm{high-$\mT$} region enriched by $\PW$+jets is used to evaluate the $\PW$+jets background in the $I_\text{rel}(\ell)<0.15$ isolation regions from the ratio between \textrm{high-$\mT$} and \textrm{low-$\mT$}. 

The number of multijet background events (QCD) in the signal region is given by :
\begin{linenomath*}\begin{equation}
N^{\text{QCD}}_{\text{low-}\mT}(A)= N^{\text{QCD}}_{\text{low-}\mT}(B)\  f^{\text{OS/SS}}_{\text{low-}\mT}.
\end{equation}\end{linenomath*}
Here $f^{\text{OS/SS}}_{\text{low-}\mT}$ is the ratio of the number of events for data in the $C$ and $D$ regions where non-QCD events were subtracted.
$N^{\text{QCD}}_{\text{low-}\mT}(B)$ is the number of events $N_{\text{low-}\mT}(B)$ observed in region $B$, where EWK and  $\PW$+jets backgrounds were subtracted: 
\begin{linenomath*}\begin{equation}
N^{\text{QCD}}_{\text{low-}\mT}(B)=N_{\text{low-}\mT}(B)-N^{\text{EWK}}_{\text{low-}\mT}(B)-N^\PW_{\text{low-}\mT}(B).
\end{equation}\end{linenomath*}
$N^\PW_{\text{low-}\mT}(B)$ is obtained from the \text{high-\mT} region : 
\begin{linenomath*}\begin{equation}
N^\PW_{\text{low-}\mT}(B)=N^\PW_{\text{high-}\mT}(B)\  f^{\text{SS}}_{\text{low-}\mT/\text{high-}\mT}. 
\end{equation}\end{linenomath*}
The high-\mT region is enriched in $\PW$+jets events, EWK background estimates obtained from simulation are
nevertheless subtracted. The ratio $f^{\text{SS}}_{\text{low-}\mT/\text{high-}\mT}$ is obtained from simulated $\PW$+jets events.

Similarly, the $\PW$+jets yield in the signal region $A$ is given by :
\begin{linenomath*}\begin{equation}
N^\PW_{\text{low-}\mT}(A)=N^\PW_{\text{high-}\mT}(A)\  f^{OS}_{\text{low-}\mT/\text{high-}\mT}.
 \end{equation}\end{linenomath*}
To limit statistical fluctuations relaxed criteria on the muon and $\tauhad$
isolations ($I_\text{rel}(\ell)<0.3$ and medium isolation for $\tauhad$) are used to obtain $f^{OS}_{\text{low-}\mT/\text{high-}\mT}$.

\subsection{The \texorpdfstring{\hadhad}{tauh tauh} final state}\label{sec_bkghh}
The multijet background estimation differs for the \hadhad final state since the $\PW$+jets background is negligible for this final state. A sideband region is defined from which the yields and shapes
are obtained. This sideband region is the same as the signal region except that the isolation requirements of the two \tauhad candidates are 
relaxed, $I_\text{rel}(\ell)<0.15$ for a \textrm{tight} and $I_\text{rel}(\ell)>0.15$ for a \textrm{loose} selection. 
The purpose of relaxing the criteria for both \tauleptons, and not only for one, is to gain in statistical precision. 
The extrapolation from the sideband region to the signal region to estimate the QCD contribution $N_\text{QCD}$ is obtained by applying 
a so called \textrm{loose-to-tight} isolation scale factor, denoted $f^\text{SS}_\text{tight/loose}$:
\begin{linenomath*}\begin{equation}
N^\text{QCD}=f^\text{SS}_\text{tight/loose}N^\text{OS}_\text{loose}.
\end{equation}\end{linenomath*}
This scale factor is computed in SS events, $f^\text{SS}_\text{tight/loose}= N^\text{SS}_\text{tight}/N^\text{SS}_\text{loose}$. 
The number of events measured in the three regions, namely
$N^\text{OS}_\text{loose}$, $N^\text{SS}_\text{tight}$, and $N^\text{SS}_\text{loose}$ is evaluated by removing the residual EWK backgrounds estimated by a MC simulation.

\section{Tau lepton spin observables} 
\label{section_optimal_observables}

\subsection{Decay angles}

The spin of the \taulepton is transferred into the total angular momentum of its decay products. Therefore, the orientation of the spin of the \taulepton can be revealed from the 
angular distributions of the decay products relative to the direction of the \taulepton momentum and relative to each other.

\begin{figure}[hbtp]
\centering
\includegraphics[width=0.3\textwidth]{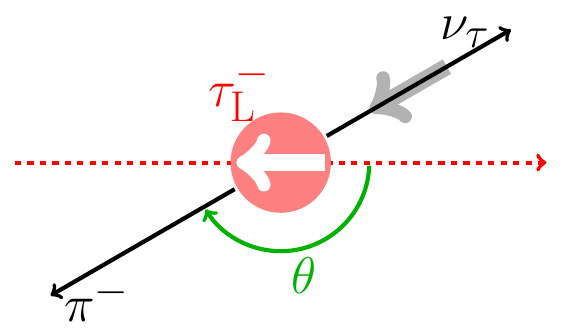}
\hfill
\includegraphics[width=0.3\textwidth]{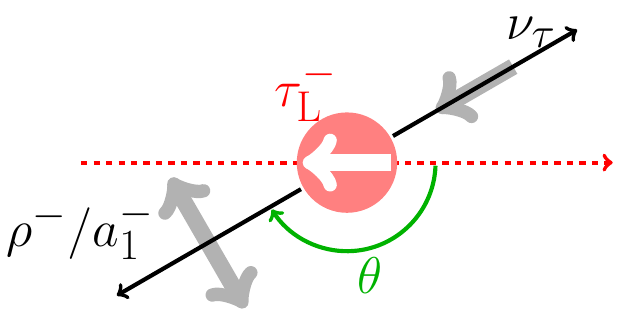}
\hfill
\includegraphics[width=0.3\textwidth]{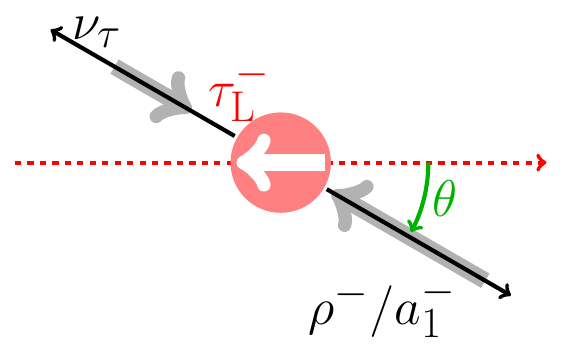}
\\
\includegraphics[width=0.3\textwidth]{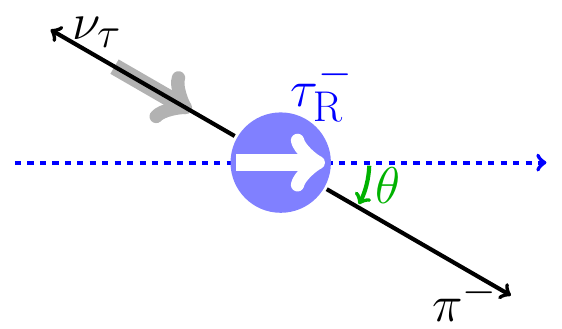}
\hfill
\includegraphics[width=0.3\textwidth]{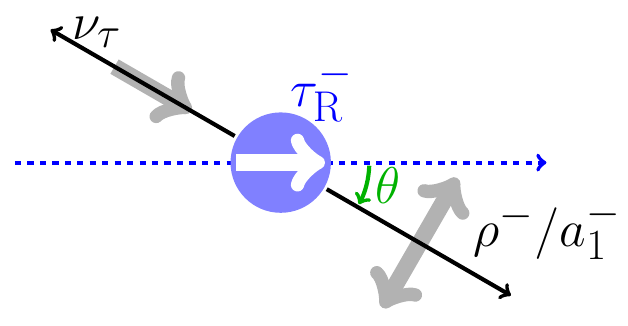} 
\hfill
\includegraphics[width=0.3\textwidth]{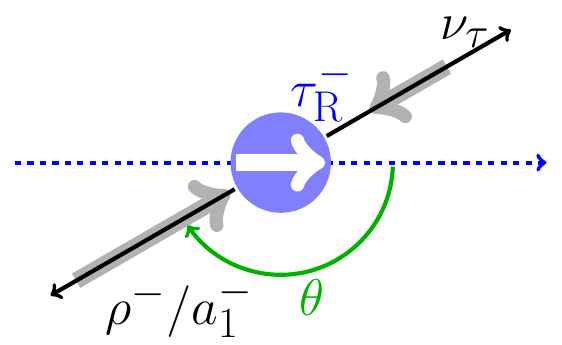}

\caption{Definition of the angle $\theta$ in the \tauleptonminus rest frame for the decays $\PGtm\to\Ph^-\PGn \, (\Ph^-=\PGpm,\,\PGrm,\,\PaDom)$,    
 upper row  for left-handed \taulepton \tauleftminus, lower row for right-handed \taulepton \taurightminus.  The thick arrows indicate the spin directions of the particles. }
\label{figure_theta_tau_to_pi_v}
\end{figure}

Figure~\ref{figure_theta_tau_to_pi_v} illustrates the relation between the \tauleptonminus helicity and the polarization of the decay products.
The first angle to consider is the angle $\theta$ between the boost direction of the \taulepton projected 
onto the \taulepton rest frame and the momentum direction of the hadron. 
\begin{linenomath*}\begin{equation}\label{theta_angle_eq}
\cos\theta = \vec{n}_{\PGt}\cdot \vec{n}_{\Ph^{\pm}} = 2 \frac{E(\Ph^{\pm})}{E(\tau^{\pm})} - 1, 
\end{equation}\end{linenomath*}
where 
$\vec{n}_{\PGt}$ and  
$\vec{n}_{\Ph^{\pm}}$ are unit vectors in the \taulepton rest frame pointing in the \PGt boost direction and along the hadron momentum, respectively.
The quantities $E(\Ph^{\pm})$ and $E(\PGt^{\pm)}$ are the energies of the hadron and \taulepton in the laboratory frame, respectively.
In the case of a leptonic \tauleptonminus decay, the vector of the hadron is replaced by the one of the lepton, but two neutrinos are emitted instead of one, leading to a more complicated situation. 
In the experimentally favored case where the charged lepton takes most of the \tauleptonminus energy, the neutrino and anti-neutrino are emitted collinear and the pair is opposite to the charged lepton, and their spins will add up to zero.  
The charged lepton will carry the spin of the \tauleptonminus, and it is preferentially emitted against the direction of \tauleptonminus spin.

In the decay 
$\PGtm\to\PGpm\PGn$,  
there is zero orbital momentum in the two-body \tauleptonminus decay and, since the pion carries no spin,  
angular momentum conservation requires the neutrino to carry the spin of the \tauleptonminus. 
In the \tauleptonminus rest frame, the neutrino and pion are emitted back-to-back.  Since the
neutrino is always left-handed, the $ \PGpm$ prefers small values for the angle $\theta$ in the decay of a right-handed \tauleptonminus (\taurightminus)  and large values in case the \tauleptonminus is left-handed (\tauleftminus). 
In this case, the angle $\theta$ carries full spin information. 

The decay into a spin-one resonance, \PGr or $\PaDo$,  also
offers the simplicity of a two-body decay, like the $\PGtm\to\PGpm\PGn$ decay, but with 
more complicated dynamics since the \PGr and $\PaDo$ resonances can have longitudinal and transverse polarization. 
Conservation of angular momentum allows the \PGr and $\PaDo$ helicities to be equal to $\lambda_{\mathrm{V}}= 0$ or $-1$.   

If the \tauleptonminus is in the right-handed state, the $\mathrm{V}$ ($\mathrm{V}=\PGr, \PaDo$) resonance tends to be in a longitudinally polarized state ($\lambda_{\mathrm{V}} =0$). 
Conversely, if the \tauleptonminus is left-handed, 
the $\mathrm{V}$ is preferably transversely polarized ($\lambda_{\mathrm{V}} = -1$ ).  Combining the spin amplitudes for all possible configurations of the $\mathrm{V}$ resonance and \tauleptonminus helicities, one gets~\cite{Tsai:1971vv}:
\begin{linenomath*}\begin{equation} \label{combinedampl}
\frac{1}{\Gamma}\frac{\rd\Gamma}{\rd\cos\theta} \propto 1 + \alpha_{\mathrm{V}}\lambda_{\PGt}\cos\theta,
\end{equation}\end{linenomath*}
where the dilution factor $\alpha_{\mathrm{V}} = (\abs{\mathcal{M}_{\mathrm{L}}}^{2} - \abs{\mathcal{M}_{\mathrm{T}}}^{2})/(\abs{\mathcal{M}_{\mathrm{T}}}^{2} + \abs{\mathcal{M}_{\mathrm{L}}}^{2}) = (m^{2}_{\PGt} - 2\unit{m}^{2}_{\mathrm{V}})/(m^{2}_{\PGt} + 2\unit{m}^{2}_{\mathrm{V}})$  is a 
result of the presence of the transverse amplitude $\mathcal{M}_{\mathrm{T}}$ of the $\mathrm{V}$ resonance in addition to the longitudinal amplitude $\mathcal{M}_{\mathrm{L}}$. The value of the factor $\alpha_{\mathrm{V}}$ characterizes the sensitivity of the $\cos\theta$ observable. 
For comparison, in the \tauleptonminus decay to the $\PaDo$~resonance, $\alpha_{\PaDo} = 0.021$, in the  \PGr meson decay, $\alpha_{\PGr}= 0.46$  and in the  
pion decay, $\alpha_{\PGp} = 1$. Consequently, the sensitivity to the \tauleptonminus helicity in $\tau\to\mathrm{V} \PGn$ decays is strongly reduced if only the 
angle $\theta$ is considered. 

The spin of $\mathrm{V}$ is transformed into the total angular momentum of the decay products and thus can be retrieved by analyzing the subsequent $\mathrm{V}$ decay.

\begin{figure}[htb]
\centering
\includegraphics[width=0.8\textwidth]{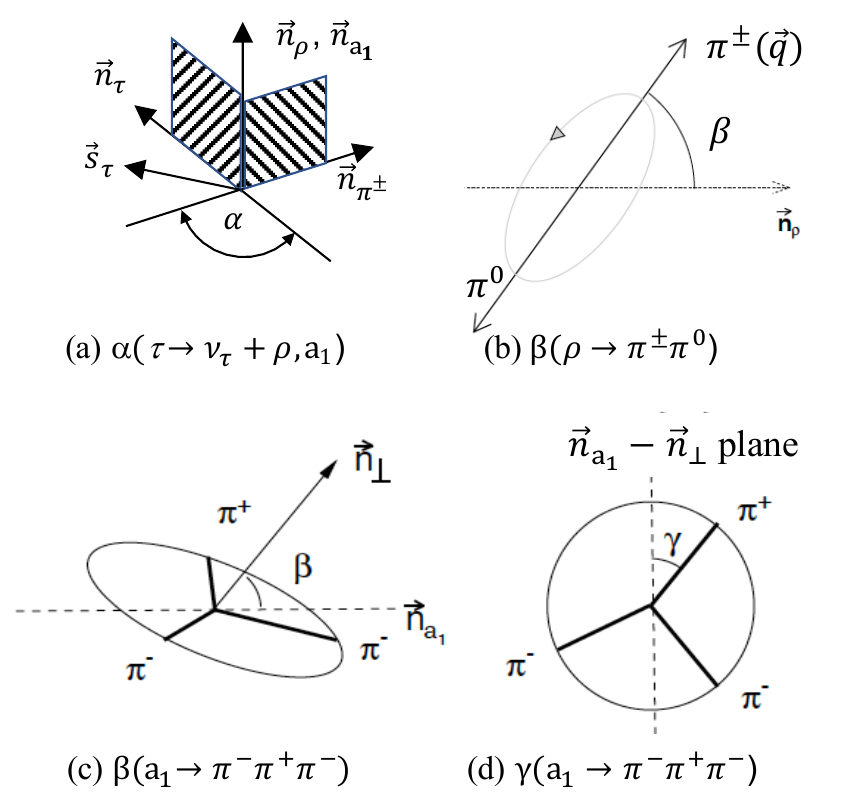}
\caption{Definitions of (a) the angle $\alpha$ in both  $\PGtm\to\PGrm\PGn$ and  $\PGtm\to\Pa^{-}_1\PGn$,  (b) the angle $\beta$ in $\PGtm\to\PGrm\PGn, \PGrm\to\PGpm\PGpz$ and (c) in $\PGtm\to\Pa^{-}_1\PGn, \Pa^{-}_1\to\PGpm\PGpp\PGpm$, and finally (d) the angle $\gamma$ for the decay of $\Pa^{-}_1\to\PGpm\PGpp\PGpm$. Figures b, c, and d have been taken and refurbished from Ref.~\cite{Cherepanov:2018wop}.}
\label{figure_all_angles}
\end{figure}

There are two additional angles to be considered in the decay of $\PGtm\to\PGrm\PGn $ followed by $\PGrm\to\PGpm\PGpz$: 
The angle $\beta$ denotes the angle between the direction of the charged pion in the \PGr meson rest frame and the direction of the \PGr meson momentum, given by:
\begin{linenomath*}\begin{equation} \label{betaequ}
\cos\beta = \vec{n}_{\Ph^\pm} \cdot \vec{n}_{\PGr} = \frac{m_{\PGr}}{\sqrt{m_{\PGr}^2-4m_{\PGp}^2}} \frac{E_{\PGpm}-E_{\PGpz}}{|p_{\PGpm}-p_{\PGpz}|}, 
\end{equation}\end{linenomath*}
where   
$\vec{n}_{\Ph^\pm}$  
is a unit vector along the direction of the charged pion in the \PGr meson rest frame and  
$\vec{n}_{\PGr}$  
is the direction of the momentum of the \PGr projected onto its rest frame.
The rotation plane of two pions is aligned correspondingly to the 
spin of the \PGr resonance. In the case of $\lambda_{\PGr} = 0$,  the angle $\beta$ tends to small values, and to large values if $\lambda_{\PGr} = -1$. The angle $\beta$ in $\PGtm\to\PGrm\PGn $ is sketched in Fig.~\ref{figure_all_angles}b. 
The expression~\ref{betaequ} for this angle contains only quantities that can be measured directly and is therefore a very powerful variable to discriminate left- and right-handed helicity states.

A similar, but not identical, angle $\beta$ is constructed for the $\PGtm\to\Pa^{-}_1\PGn$. In this case $\beta$ is defined as the angle between the normal to the $3\PGp$ decay plane and the $\PaDo$ meson momentum vector projected onto the  $\PaDo$ meson rest frame. The angle is shown in Fig.~\ref{figure_all_angles}c. 

A further angle  $\alpha$ is defined by the two planes spanned by vectors 
($\vec{n}_{\mathrm{V}}, \vec{n}_{\PGt}$) and ($\vec{n}_{\mathrm{V}}, \vec{n}_{\Ph^\pm}$), respectively:
\begin{linenomath*}\begin{equation} \label{alpharhoeq}
\cos\alpha = \frac{(\vec{n}_{\mathrm{V}}\times\vec{n}_{\PGt})\cdot(\vec{n}_{\mathrm{V}}\times\vec{n}_{\Ph^\pm})} {\abs{\vec{n}_{\mathrm{V}}\times\vec{n}_{\PGt}}\abs{\vec{n}_{\mathrm{V}}\times\vec{n}_{\Ph^\pm}\ }}.
\end{equation}\end{linenomath*}
Here, all vectors are defined in the resonance rest frame.  The vector $\vec{n}_{\mathrm{V}}$  
denotes the boost direction of the \PGr or $\PaDo$ resonance projected onto its rest frame, and  
$\vec{ n}_{\PGt}$ is the boost direction of the \tauleptonminus in the rest frame of the meson.  
The vector 
$\vec{n}_{\Ph^\pm}$ denotes the momentum direction of the $\PGpm$ in case of $\PGtm\to\PGrm\PGn$ decay and 
 $\PGpp$ in $\PGtm\to\Pa^{-}_1\PGn$ decay. 
The angle $\alpha$ describes the correlation between the helicity of the 
\taulepton and the decay products of the \PGr or $\PaDo$ resonance. The angle $\alpha$ is illustrated in Fig.~\ref{figure_all_angles}a.

An additional angle $\gamma$ can be introduced in the $\PGtm\to\Pa^{-}_1\PGn$ decay; the angle  describes the relative orientation of the pions in the decay plane.
This angle is illustrated also in Fig.~\ref{figure_all_angles}d. 
All spin sensitive angles $\alpha, \beta, \gamma$ and $\theta$ can be combined into a unique one-dimensional variable, without loss of sensitivity, as described in the next section.

\subsection{Optimal observables}\label{optimal_observalbe_section}

Optimal variables to measure the helicity of \tauleptonsminus have been widely explored at LEP and are explained in Ref.~\cite{Davier:1992nw}. 
The idea is to replace the multidimensional fits over all relevant decay angles by a one dimensional fit to a unique variable $\omega(\theta, \alpha, \beta, \gamma, ... ) $, which can be derived from the squared matrix elements of the decay for positive and negative helicity. 
This procedure is equivalent to the use of the concept of the polarimetric vector.

In general, the differential width of any decay of a polarized \tauleptonminus can be described using the polarimetric vector and has the simplified form~\cite{Tsai:1971vv,Jadach:1993hs,Cherepanov:2018wop}:
\begin{linenomath*}\begin{equation}\label{decaydistr12}
\rd\Gamma = \frac{\abs{\overline{\mathcal{M}}}^{2}}{2m_{\PGt}}(1-h_{\mu}s^{\mu})\rd S,
\end{equation}\end{linenomath*}
where $\rd S$ is the element of Lorentz-invariant phase space, $s$ is the four-vector of the spin of the \tauleptonminus and $h$ is the polarimetric four-vector.
In the rest frame of the \tauleptonminus, assuming that the \tauleptonminus spin is aligned with the \tauleptonminus momentum (all transverse $\PGtm$ spin components are zero), this reads:
\begin{linenomath*}\begin{equation}\label{decaydistr2}
\frac{\rd\Gamma}{\rd\cos\zeta_{h}} \propto \frac{1}{2}(1 + \mathcal{P}_{\PGt}\cos\zeta_{h}),
\end{equation}\end{linenomath*}
where $\mathcal{P}_{\PGt}$ is the \tauleptonminus polarization 
for a given \tauleptonminus decay and 
$\zeta_{h}$ is the angle between the space-like 
polarimetric vector $\vec{h}_{\PGt}$ and $\vec{n}_{\PGt}$, the direction of the \tauleptonminus as defined above. 

The polarimetric vector $\vec{h}_{\PGt}$ 
depends on the particular final state. This vector describes 
how the tau decay products propagate depending on the tau spin;
for \tauleptonminus decays into hadrons it depends on 
the momenta of all the hadrons in the decay chain and the decay mode of the intermediate resonance.

In the case of $\PGtm\to\PGpm\PGn$, the polarimetric vector is aligned with the direction of the $\PGpm$ meson:  $\vec{h}_{\PGt} = \vec{n}_{\PGp}$, where  $\vec{n}_{\PGp}$ is a unit vector pointing along the momentum of the pion.

The situation in $\PGtm\to(\PGrm\to\PGpm\PGpz)\PGn$ and $\PGtm\to(\PaDom\to\PGpm\PGpp\PGpm)\PGn$ decays is more complicated due to 
the resonance structure. The general expressions for
polarimetric vectors in the decays $\PGtm\to\PGrm\PGn$  and $\PGtm\to\PaDom\PGn$ is reported in Ref.~\cite{Cherepanov:2018wop} and references therein, including a discussion of the hadronic form factor in $\PGtm\to\PaDom\PGn$ decays. 

We denote the optimal observable for a single \tauleptonminus decay as $\omega(\Ph)$, which is defined in Ref.~\cite{Davier:1992nw}:
\begin{linenomath*}\begin{equation}\label{omega_opt}  
\omega(\Ph)= \cos\zeta_h.
\end{equation}\end{linenomath*}

The optimal observable $\omega(\Ph)$ carries the full sensitivity and has the same unique distribution distinguishing left and right \taulepton helicities for all hadronic decay modes.
The simple linear distribution of $\omega(\Ph)$ (for $\Ph=\PGp,\PGr,\PaDo$) is shown in Fig.~5 of Ref.~\cite{Cherepanov:2018wop}. The use of such an optimal
variable simplifies the extraction of the polarization and combines all spin
sensitive angles, even those with low discrimination power, into a one-dimensional quantity.

 For the  $\PGtm\to\PaDom\PGn$ decay, we use the parametrization provided by the CLEO
Collaboration~\cite{CLEO:1999rzk} for the hadronic form factor,
which includes in total seven resonances that describe the intermediate states of the $\PaDom\to\PGpm\PGpp\PGpm$ decay. The CLEO parametrization for this decay
mode is also used in the \TAUOLA~\cite{Jadach:1993hs, Davidson:2010rw} and \PYTHIA8~\cite{Sjostrand:2014zea} Monte Carlo generators.

\subsection{Helicity correlations}

{\tolerance=800
The sensitivity to the polarization measurement can be improved by exploiting the anti-correlation of helicity within a \taulepton pair. 
The method, proposed in Ref.~\cite{Davier:1992nw}, defines a new optimal observable for the \taulepton pair. Denoting $\omega_1$ 
and $\omega_{2}$ to be the observables for $\PGt_{1}$ and $\PGt_{2}$, one can define the variable:
\begin{linenomath*}\begin{equation}
\Omega=\frac{\omega_1+\omega_{2}}{1+\omega_1\omega_2}.
\end{equation}\end{linenomath*}
Similarly to the optimal observable for a single \PGt decay, the distribution of $\Omega$ for a \PGt pair is identical for all hadronic decay modes, as shown in Fig.~6 in Ref.~\cite{Cherepanov:2018wop}.
\par}

 \subsection{Reconstruction of \texorpdfstring{$\omega(\Ph)$}{omega[h]} and the choice of discriminators}
The performance of the kinematical reconstruction of \taulepton decays and the polarization observables
strongly depends on the decay mode.
Therefore, sensitivity is improved by treating separately each decay mode, which results
in 11 categories in the \elmu, \elhad, \muhad, and \hadhad channels. 
These categories are summarized in the first two columns of
Table~\ref{table_disciminators_per_event_categories}. 
In the fully hadronic channel, the three combinations containing $\PGtm\to\PGrm\PGn$ events are combined into one category for the choice of the optimal discriminator.

\begin{table}[hbp]
\topcaption{Final choice of discriminators in the various event categories}
\centering
\begin{tabular}{llll}
\hline
Channel & Category           & \multicolumn{2}{l}{Discriminator} \\
\hline
\elmu            & \emCombinedOneprongOneprong & $m_\text{vis}(\Pe,\PGm)$     & visible mass \\[\cmsTabSkip]
\elhad           & \etCombinedAoneOneprong     & $\omega(\PaDo)$       & optimal observable with \textsc{SVfit} \\
                 & \etCombinedRhoOneprong      & $\omega_\text{vis}(\PGr)$       & visible optimal observable \\
                 & \etCombinedOneprongOneprong & $\omega(\PGp)$       & optimal observable with \textsc{SVfit}  \\[\cmsTabSkip]
\muhad           & \mtCombinedAoneOneprong     & $\omega(\PaDo)$       & optimal observable with \textsc{SVfit} \\
                 & \mtCombinedRhoOneprong      & $\omega_\text{vis}(\PGr)$ & visible optimal observable \\
                 & \mtCombinedOneprongOneprong & $\omega(\PGp)$      & optimal observable with \textsc{SVfit} \\[\cmsTabSkip]
\hadhad          & \ttCombinedAoneAone         & $m_\text{vis}(\PaDo,\PaDo)$   & visible mass \\
                 & \ttCombinedAoneOneprong     & $\Omega(\PaDo,\PGp)$   & combined optimal observable with \textsc{SVfit} \\
                 & \ttRho                      & $\omega_\text{vis}(\PGr)$ & visible optimal observable (for leading $\PGr$) \\
                 & \ttCombinedOneprongOneprong & $m_\text{vis}(\PGp,\PGp)$   & visible mass \\
\hline
\end{tabular}
\label{table_disciminators_per_event_categories}
\end{table}
The reconstruction of the optimal observables $\omega(\Ph)$ requires the knowledge of the rest frame of the \taulepton or of 
both \tauleptons for the calculation of the quantity $\Omega$. 
Because of the emission of the neutrino, the momentum direction of \tauleptons can be measured directly only in cases where the decay vertex can be reconstructed, for example for decays like \tauthreepi. In most other cases, the \textsc{SVfit}~\cite{Bianchini:2016yrt} 
algorithm for the reconstruction of the \taulepton decay is used.  
The \textsc{SVfit} algorithm has been modified for the purpose of this analysis, such that: 
\begin{itemize}
\item the four-vector of individual \PGt leptons is calculated by \textsc{SVfit};
\item the mass of the di-$\PGt$ system is constrained to the \PZ  boson mass to improve the resolution on the four-vectors of the two \PGt leptons;
\end{itemize}
The reconstructed \tauleptonminus four-vectors are used to boost the four-vectors of the $\PGtm$ decay products to the \tauleptonminus rest frame and to compute 
the optimal observable $\omega(\Ph)$ in the \tauleptonminus rest frame.

Since the performance of the \textsc{SVfit} algorithm is limited, in some event categories a better sensitivity is often obtained from observables that are 
measured directly in the detector, \eg, the angle $\beta$ in \PGr decays, and the angles $\beta$ and $\gamma$ in $\PGtm\to\PaDom\PGn$ decays, as defined by Eq.~(\ref{betaequ}). 
These quantities are denoted hereafter in the text as visible observables
$\omega_\text{vis}$ or $\Omega_\text{vis}$.  

In addition, it has been shown in Ref.~\cite{Alemany:1991ki} that the energies of the \taulepton decay products and the acollinearity angle between the decay 
products of the two \tauleptons in the laboratory frame also carry information about the helicity states of the \tauleptons.  
A suitable observable, which includes both quantities, is the invariant mass of the visible decay products of both \tauleptons
\begin{linenomath*}\begin{equation}
m_\text{vis}=\sqrt{2 E_{1}E_{2}(1-\cos\varphi({\tau^\text{vis}_{1}, \tau^\text{vis}_{2}}))},
\end{equation}\end{linenomath*}
where $E_{1}$ and $E_{2}$ are the energies of the detected leptons and hadrons in the final states of $\PGt_{1}$ and $\PGt_{2}$ and $\varphi({\tau^\text{vis}_{1}, \tau^\text{vis}_{2}})$ the 
acollinearity angle defined as the angle between the decay products of the two \tauleptons.
The visible mass $m_\text{vis}$  is used for $\Pepm+\PGm^{\mp}$, $\PGpp+\PGpm$, and $\Pa^{+}_1+\Pa^{-}_1$  pairs.

The decision of which observable to use in each category was made on the basis of a likelihood scan of the expected sensitivity in each category with the simulated data sample.  All possible observables were evaluated for each category. 
The result of the best choice is given in the last column of Table~\ref{table_disciminators_per_event_categories}. 
For the category $(\PGr+\tauhad)$ we use $\omega_\text{vis}(\PGr)=\cos\beta$ for all possible decays of \tauhad, and
for the combination where both \tauleptons decay via a \PGr we use $\omega_\text{vis}(\PGr)$ of the \PGr with the highest $p_T$.

\section{Systematic uncertainties} 
\label{sec_syst}

The estimated systematic uncertainties follow closely the analysis of Ref.~\cite{CMS:2021sdq} and an earlier analysis that uses the same data set~\cite{2018283}. 
Our uncertainties are briefly discussed in the following, indicating differences and refinements compared with previous analyses. 
Systematic uncertainties are modeled as nuisance parameters with log-normal (gaussian) probability distribution functions for normalization (shape) uncertainties in the maximum likelihood fits described in Section~\ref{results}. 
This section presents the initial uncertainties used for the nuisance parameters.
The systematic uncertainties that affect only the normalization of signal and background processes are summarized in Table~\ref{table_systematic_uncertainties_lnn}, 
and those that could alter the distributions of the \PGt polarization observables are summarized in Table~\ref{table_systematic_uncertainties_shape}.

\begin{table}[htbp]  
\topcaption{Systematic uncertainties affecting only the normalization of templates. The table lists the estimated inital uncertainties used in the fit for these nuisance parameters. The label \textit{correlated} means that these uncertainties are common to the respective channels.}
\centering
\begin{tabular}{lcccc}
\hline
Source of uncertainty  & \multicolumn{4}{c}{Initial uncertainties per channel} \\
                                & ${\hadhad}$ & ${\muhad}$ & ${\elhad}$ & ${\elmu}$ \\
\hline
Integrated luminosity                       & 1.2\% & 1.2\%& 1.2\%& 1.2\% \\ 		
DY cross section                            & 5.6\% & 5.6\%& 5.6\%& 5.6\% \\ 		
\ttbar cross section                        & 4.2\% & 4.2\%& 4.2\%& 4.2\% \\ 		
Diboson cross section                       & 5\%   & 5\%  & 5\%  & 5\% \\ 		
EWK cross sections                           & 4\%   & 4\%  & 4\%  & 4\% \\ 
$\PW$+jets cross section \& normalization     & 4\%   & 10\% & 10\% & 20\% \\ 
QCD bkg. normalization                           & 3\%   & 20\% & 20\% & 10\% \\ 
b-tagging efficiency                               & \multicolumn{4}{c}{$\leq$0.1\% except for \ttbar and VV(1--9\%) }  \\
\Pe identification efficiency (\textit{correlated})  & \NA    & \NA   & 2\%  & 2\% \\		 
\Pe tracking efficiency (\textit{correlated})        & \NA    & \NA   & 1\%  & 1\% \\ 		
\PGm identification efficiency (\textit{correlated})& \NA    & 2\%  & \NA   & 2\% \\ 	
\hline
\end{tabular}
\label{table_systematic_uncertainties_lnn}
\end{table}

The overall luminosity uncertainty of 1.2\%  has been determined in Ref.~\cite{CMS:2021xjt}.
 The uncertainty in the Drell--Yan cross section is a combination of the uncertainty in the theoretical cross section and PDF, $\alpha_s$, and QCD scale variations were evaluated in Ref.~\cite{2018283}. 
 The uncertainties for \ttbar (4.2\%), diboson (5\%), and EWK cross sections (4\%) were established in Ref.~\cite{2018283}.
The normalization of the $\PW$+jets cross section is different for various channels due to the different contributions and methods; 
in  the \elmu channel a MC simulation was used, whereas in the \elhad and \muhad channels the $ABCD$ method was used as described in Section~\ref{sec_bkg}. 
For the \hadhad channel the $\PW$+jets contribution is very small and only the cross section uncertainty is applied.
For the QCD normalization uncertainty we assign: (i) 10\% uncertainty for the \elmu channel; (ii) 20\% for the  \elhad and \muhad channels because of  an extrapolation uncertainty from the loose isolation control region; and (iii) 3\% for the \hadhad channel because we use a data driven method, see Section~\ref{sec_bkghh}. 
The \PQb quark tagging efficiency uncertainty is applied only for the \ttbar and diboson processes.

The \Pe and \PGm identification efficiency uncertainties are 2\% and identical for the channels where electrons or muons are used~\cite{2018283}. 
The muon tracking efficiency is included in the identification efficiency. 

\begin{table}[th]  
\caption{Systematic uncertainties affecting the shapes of templates. The table lists the estimated initial uncertainties used in the fit for these nuisance parameters and their dependencies on \pt, decay mode (DM) or event selection. 
The comment ''Event-dependent'' for the \ptmiss entries indicates that these corrections vary on an event-by-event basis due to the event selection. }
\centering
\begin{tabular}{lcccc}
\hline
Source of uncertainty & \multicolumn{4}{c}{Initial uncertainties per channel} \\
                               & ${\hadhad}$ & ${\muhad}$ & ${\elhad}$ & ${\elmu}$ \\
\hline
$e\to\tauhad$ misidentification rate               & \multicolumn{3}{c}{ 12\%    DM dependent   }& \NA    \\
$\mu\to\tauhad$ misidentification rate             & \multicolumn{3}{c}{ 25\%    DM dependent  }&  \NA    \\
$\text{jet}\to\tauhad$ misidentification rate    & \multicolumn{3}{c}{  $20\% \times \pt^\text{jet} /100\GeV \leq 40\%$} &  \NA \\
Electron trigger efficiency                    & \NA     & \NA     & \multicolumn{2}{c}{\pt,$\,\eta$ dependent $\leq 2\%$ }  \\
Electron momentum scale                          & \NA     & \NA     & \multicolumn{2}{c}{Event-dependent} \\
Electron to tau misid energy scale              & \NA     & \NA     & 0.8--6.6\% & \NA \\
Muon trigger efficiency                        & \NA     &\pt,$\,\eta$ dependent $\leq 2\%$& \NA     & \pt~$\eta \leq 2\%$ \\
Muon momentum scale                              & \NA     & 0.4--2.7\% & \NA     & 0.4--2.7 \% \\
Muon to tau misid momentum scale                  & \NA     & 1\% & \NA     & \NA \\
Hadronic tau momentum scale                      & \multicolumn{3}{c}{\pt \& DM dependent $\leq 2\%$ }& \NA \\
Neutral, charged hadrons energy             & 2\%   & 2\%        & 2\% & \NA \\
Tau identification efficiency                  & \multicolumn{3}{c}{\pt \& DM dependent 2--3\%} & \NA    \\
Tau trigger efficiency                         &\multicolumn{2}{c}{\pt \& DM dependent $\leq 10\%$}  & \NA & \NA    \\
Misidentified DM $\tauhad\to\Ph^\pm$               & 2.8\% & 2.8\% & 2.8\% & \NA \\
Misidentified DM $\tauhad\to\Ph^\pm \PGpz$         & 3.2\% & 3.2\% & 3.2\% & \NA \\
Misidentified DM $\tauhad\to\Ph^\pm \Ph^\mp \Ph^\pm$   & 3.7\% &  3.7\% &  3.7\% & \NA \\
Drell--Yan MC reweighting                      & \multicolumn{4}{c}{$\leq$100\% for all channels}\\
Top \pt reweighting                         & \multicolumn{4}{c}{$\leq$100\% for all channels}\\
Parton reweighting                            & \multicolumn{4}{c}{$\leq$100\% for all channels}\\
$\pt^\text{miss}$ unclustered scale                         & \multicolumn{4}{c}{Event-dependent} \\
$\pt^\text{miss}$ recoil correction                          & \multicolumn{4}{c}{Event-dependent}  \\
Limited amount  of MC events                          & \multicolumn{4}{c}{Bin-by-bin fluctuations} \\
\hline
\end{tabular}
\label{table_systematic_uncertainties_shape}
\end{table}

The uncertainties listed in Table~\ref{table_systematic_uncertainties_shape} can possibly alter the shape of observables: the electron and muon misidentifications depend on the decay modes and lead to rate uncertainties of 12 and 25\% for electron and muon misidentification, respectively; the uncertainty in the $\text{jet}\to\tauhad$ misidentification rate is \pt dependent but does not exceed 40\%~\cite{2018283}.

The uncertainties in the efficiency of the electron and muon trigger are in the order of 2\%, but slightly \pt and $\eta $ dependent.
The muon momentum scale uncertainty is specified in three bins in $\eta$ and ranges from 0.4 to 2.7\%.
The hadronic \PGt momentum scale uncertainty is \pt and decay mode dependent but less than 2\%~\cite{2018283}. In addition we added a further 2\% contribution to account for possible differences in the charged and neutral hadron energy reconstruction.

The identification efficiencies for \tauhad candidates depend on working points of the above mentioned MVA used to refine the decay mode identification and have been evaluated in two regions of \pt, below and above 40\GeV as described in Ref.~\cite{CMS:2021sdq}. Similarly the \tauhad trigger efficiencies have a \pt and MVA dependence of up to 10\%~\cite{2018283}. The same working points are used in the \muhad and \hadhad channels and are therefore correlated between these two channels. The \elhad channel uses only the electron trigger and therefore is not sensitive to the \PGt trigger uncertainty.
 
Uncertainties related to the reconstruction of hadronically decaying \tauleptons are considered to be uncorrelated between event categories for different decay modes. 
The imperfect knowledge of the decay mode of the reconstructed, hadronically decaying \tauleptons and its uncertainty is a major systematic limitation in this analysis, but
can be controlled by the shape of the \taulepton mass for a given decay channel, which includes its contribution from wrongly classified decay modes. 
Maximum likelihood fits to the \taulepton mass distribution in data obtained normalization factors for these contributions identical to the values of the simulation except for the $\PGr+X$ category, where decays containing additional, but not identified, \PGpz{s} contribute.
A shift of 3.2\% corresponding to about one standard deviation of the parameter describing this migration was necessary to make the \taulepton mass distribution agree with data for this decay mode. 
Table~\ref{table_systematic_uncertainties_shape} lists the uncertainties of the parameters describing the migration for three important decay modes. 
For the extraction of the polarization, this correction of the migration has been applied in the $\PGm+\PGr$ channel, where the visible optimal observable $\omega_\text{vis}(\PGr)$ is exploited.
The effect on the distributions of the \PGt lepton polarization observables is estimated by varying contributions from events with misidentified hadronic decay modes symmetrically by one standard deviation. 

As pointed out in Section~\ref{optimal_observalbe_section}, the polarimetric vector in $\PGtm\to\PGpm\PGn$ and $\PGtm\to\PGrm\PGn$ decays 
is determined purely by the Lorentz structure of these decays and thus contains no QCD dynamics.
The \taulepton decay to three pions and a neutrino is dominated by the $\PaDoP{1260}$ ($J^{PG} = 1^{+-}$) resonance, which decays through the intermediate
state of ($\PGr\PGp$), with mostly $\PGrP{770}$ and an admixture of $\PGrP{1450}$ at higher
masses, followed by a \rhopipi decay. 
Theoretical systematic uncertainties originating from the imperfect knowledge of the $\PaDo\to3\PGp$ 
decay substructure have been studied in Ref.~\cite{Cherepanov:2018wop}. 
The uncertainty in the \PGt lepton polarization in the $\PaDo$ channel due to the variation of
resonance parameters is  $(\Delta \mathcal{P}_{\PGt})_\text{model} = (1.41 \pm 1.37)\times 10^{-4}$,
far below any currently reachable experimental
precision and hence neglected. 

As described in Section~\ref{section_samples}, the DY MC has been reweighted in \pt and rapidity and a variation of 100\% of this correction is used to evaluate the systematic effect. The same procedure has been applied to the parton reweighting corrections described in the same section. 

The missing unclustered energy uncertainty is applied for \ttbar and diboson processes and the $\pt^\text{miss}$ recoil correction is varied within the limits determined during the computation of the correction.

A further important contribution to the overall systematic uncertainty arises from the limited amount of MC data, which contributes to bin-to-bin fluctuations.

\section{Measurement of the \texorpdfstring{\taulepton}{tau-lepton} polarization} \label{section_polarisation_measurement}

The average \taulepton polarization \averagepolarisation is defined as
\begin{linenomath*}\begin{equation}
 \label{equation_definition_polarisation}
\averagepolarisation = \frac{N(\ztautaupospol) - N(\ztautaunegpol)}{N(\ztautaupospol) + N(\ztautaunegpol)},
\end{equation}\end{linenomath*}
where $N(\ztautaupospol)$ and $N(\ztautaunegpol)$ are the numbers of \PZ boson decays into \tauleptons with positive and negative $\PGtm$ lepton helicity, respectively. 
The average polarization depends on
the accepted mass range of the \PGt lepton pair.
The numbers \ztautauposnegpol are obtained by a maximum likelihood fit of representative templates to the distributions of the observables listed in 
Table~\ref{table_disciminators_per_event_categories}. The signal templates are derived from simulated signal events. 
The background templates are composed of the contributions estimated from data or from simulation, as described in Section~\ref{sec_bkg}.

\begin{figure}[htb]
\centering                  
    \includegraphics[width=0.32\textwidth]{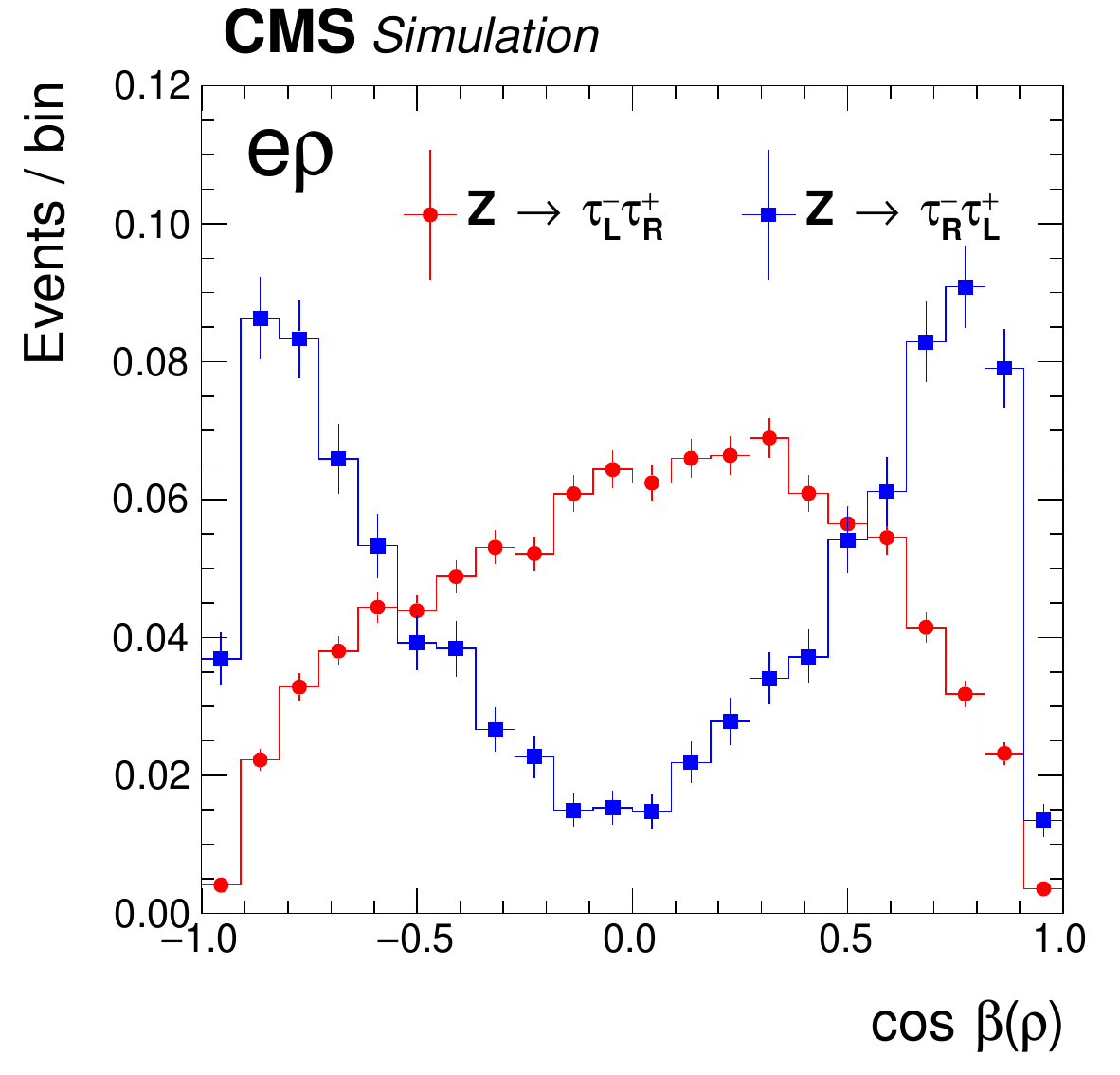}
    \includegraphics[width=0.32\textwidth]{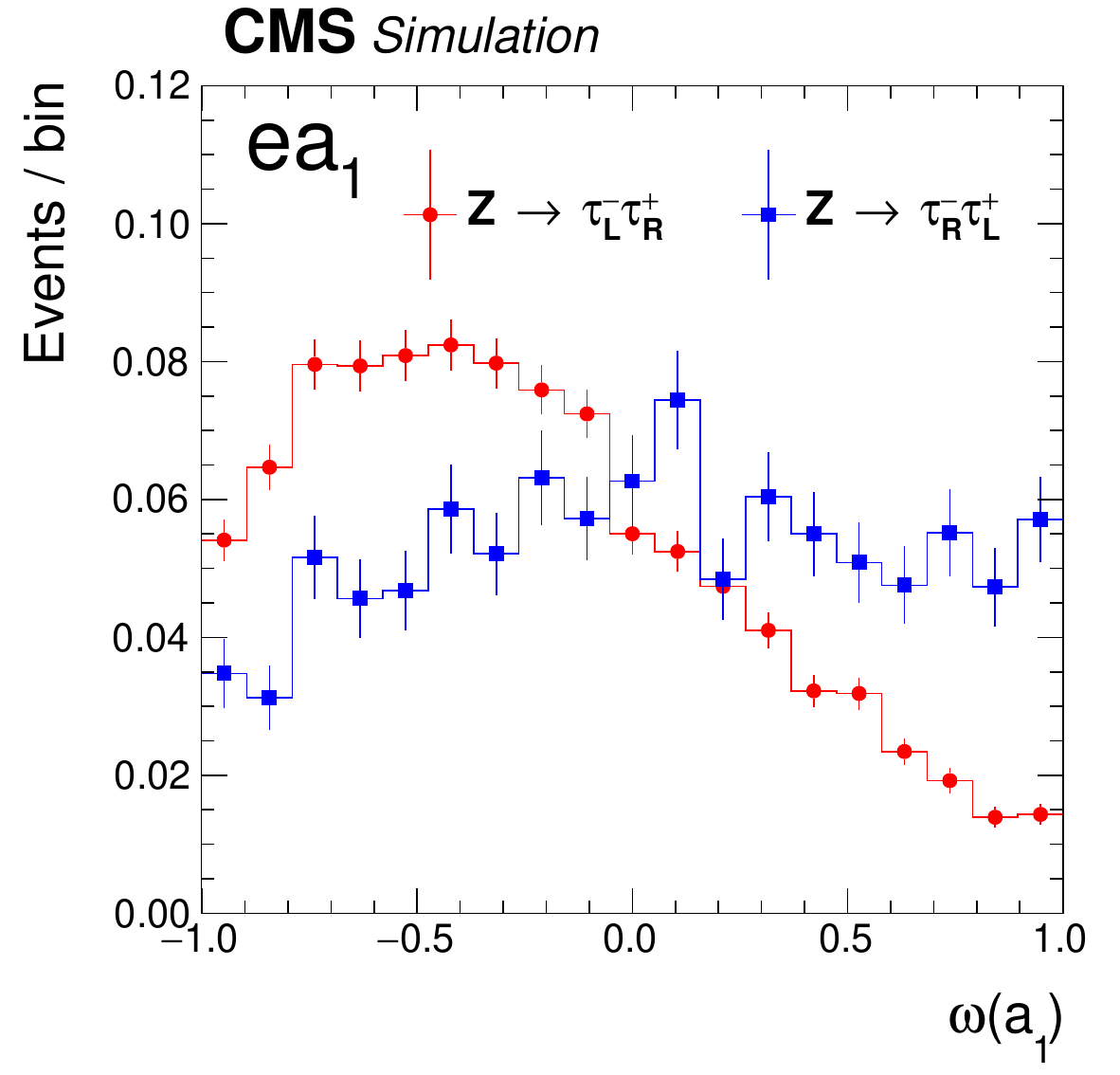}
    \includegraphics[width=0.32\textwidth]{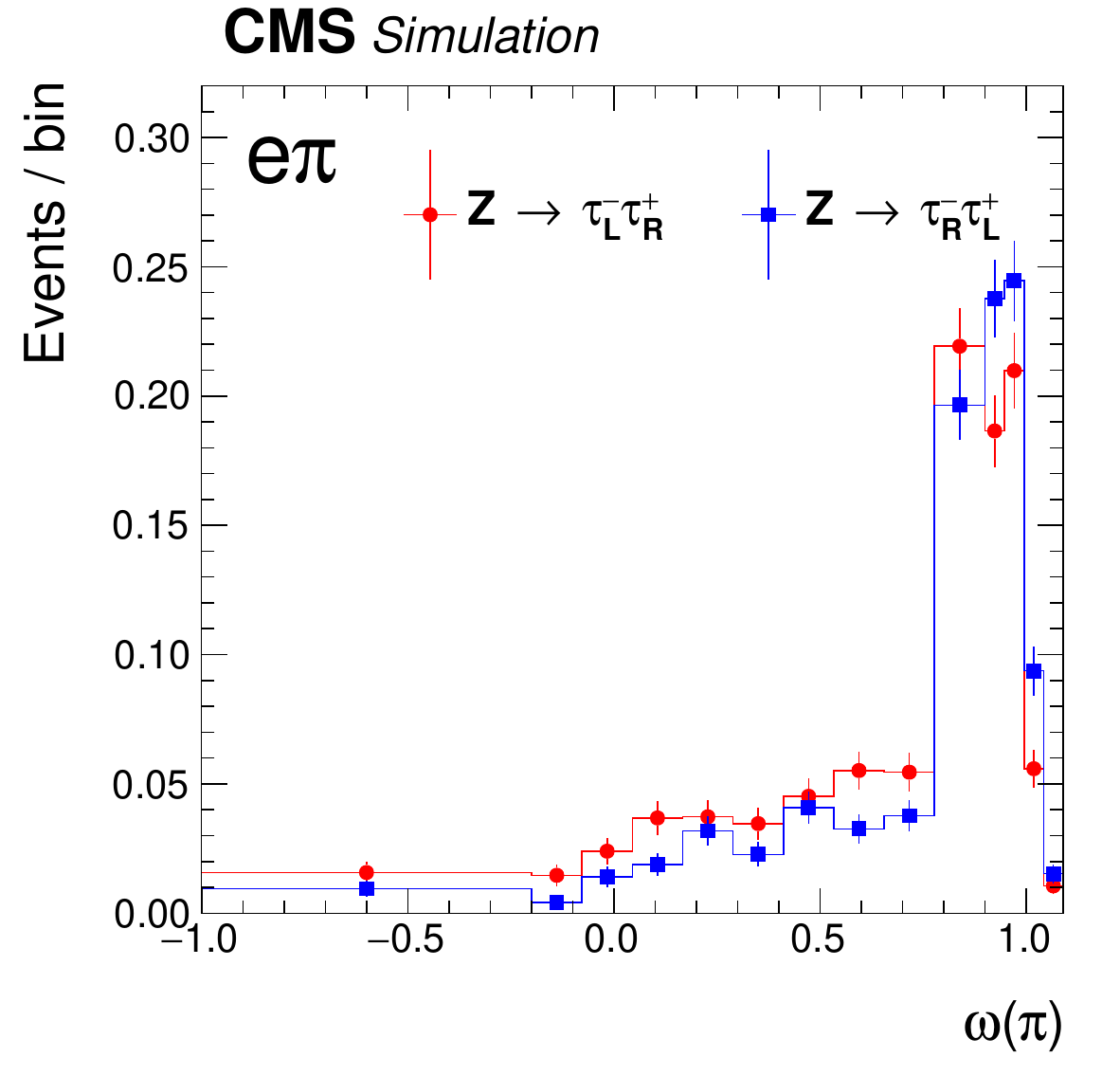}\\
    \includegraphics[width=0.32\textwidth]{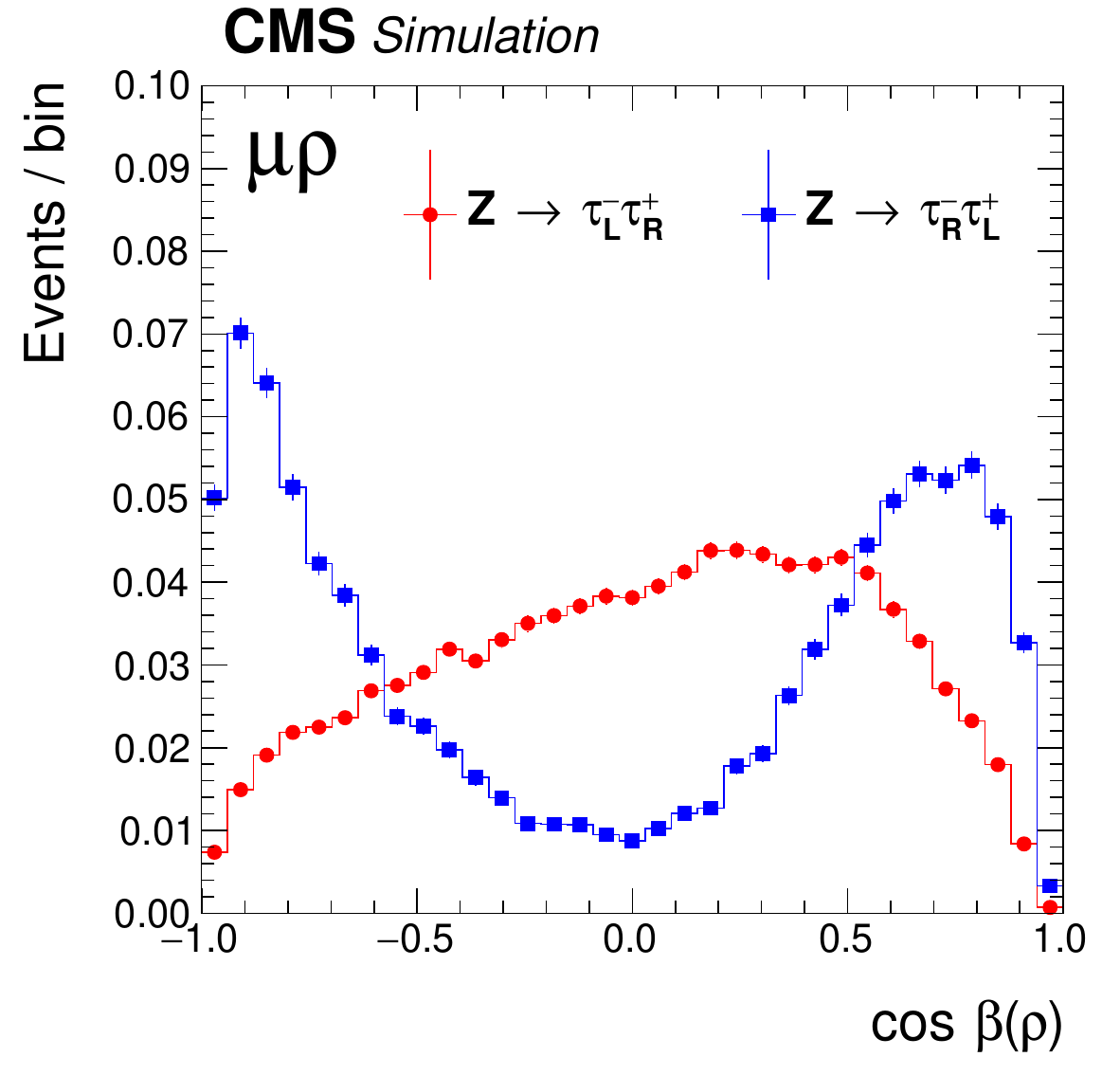}
    \includegraphics[width=0.32\textwidth]{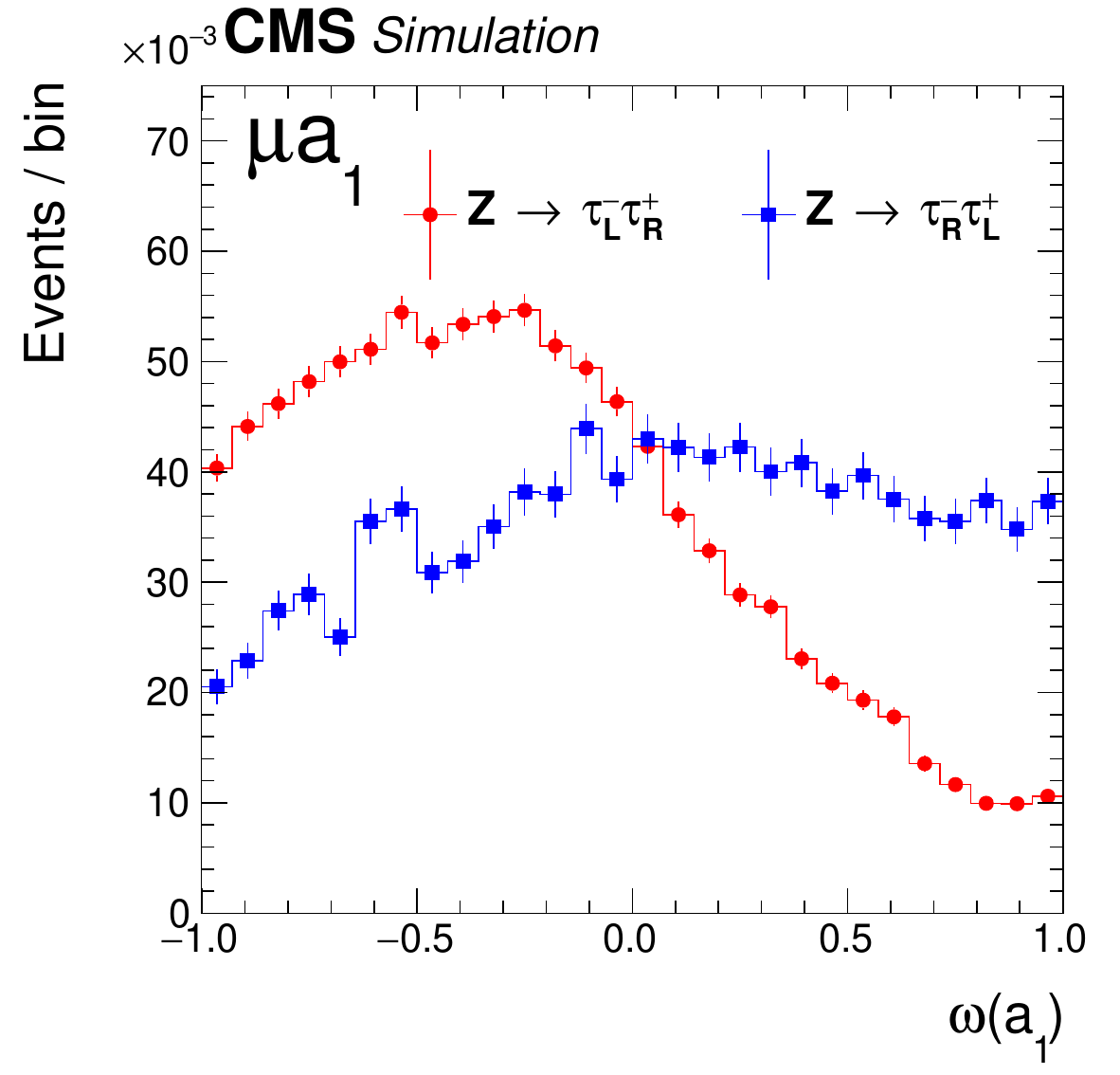}
    \includegraphics[width=0.32\textwidth]{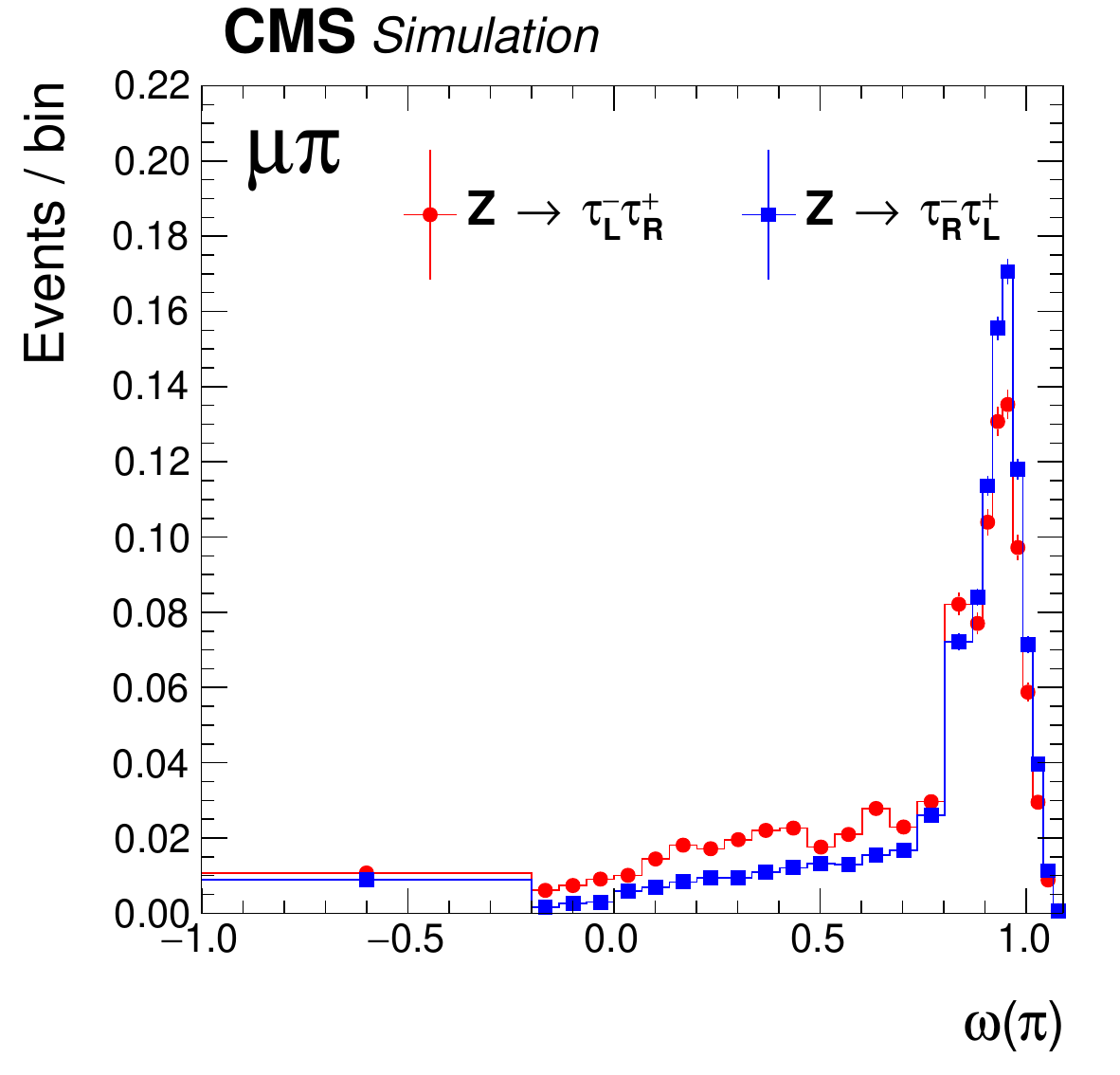}
\caption{Some examples of templates for six \tauleptonminus decays:  $\Pe+\PGr, \Pe+\PaDo, \Pe+\PGp$ in the upper row and $\PGm+\PGr, \PGm+\PaDo, \PGm+\PGp $  in the lower row. The blue and red lines indicate right and left-handed \tauleptonsminus, respectively. The templates have been rescaled to correspond to zero polarization at the generator level. The uncertainty bars are statistical only and correspond to the limited MC samples after all selections. The shapes of the templates depend on the decay mode and on the nature of the chosen discriminating observables. }
\label{fig_template}
\end{figure}

The extracted numbers from these maximum likelihood fits and thus the polarization are always an average over a range of intrinsic polarization given by  
\begin{linenomath*}\begin{equation}
\label{pol_i}
\polarisation = \polarisation \left( \sintwoth, \hat{s}, \PQq \right),
\end{equation}\end{linenomath*}
which depends on the effective weak mixing angle,  the quark-antiquark center-of-mass energy $\sqrt{\hat{s}}$ and the quark type \PQq specifying whether the \PZ boson has been produced via a pair of up- or down-type quarks.

The measured average polarization can be written as
\begin{linenomath*}\begin{equation}
\label{equation_average_polarisation_efficiency}
\left\langle \polarisation(\sintwoth) \right\rangle = \frac{\sum_{\PQq=\PQu, \PQd} \int  \rd \hat{s} \; \polarisation \left( \sintwoth, \hat{s}, \PQq \right) w \left( \sintwoth, \hat{s}, \PQq \right)}{\sum_{\PQq=\PQu, \PQd} \int \rd \hat{s} \, w \left( \sintwoth, \hat{s}, \PQq \right)}.
\end{equation}\end{linenomath*}

Here $w \left( \sintwoth, \hat{s}, \PQq \right)$ denotes the weights in the averaging. These weights include two effects: (i) the dependence of the cross section $\sigma$ on the quark-antiquark squared center-of-mass energy $\hat{s}$ and the quark type \PQq for a given value of the mixing angle \sintwoth; and (ii) the acceptance of the detector and the selection efficiency in the analysis, which both depend on $\hat{s}$ and quark \PQq as well. 

\begin{figure}[th]
  \centering     
    \includegraphics[width=0.32\textwidth]{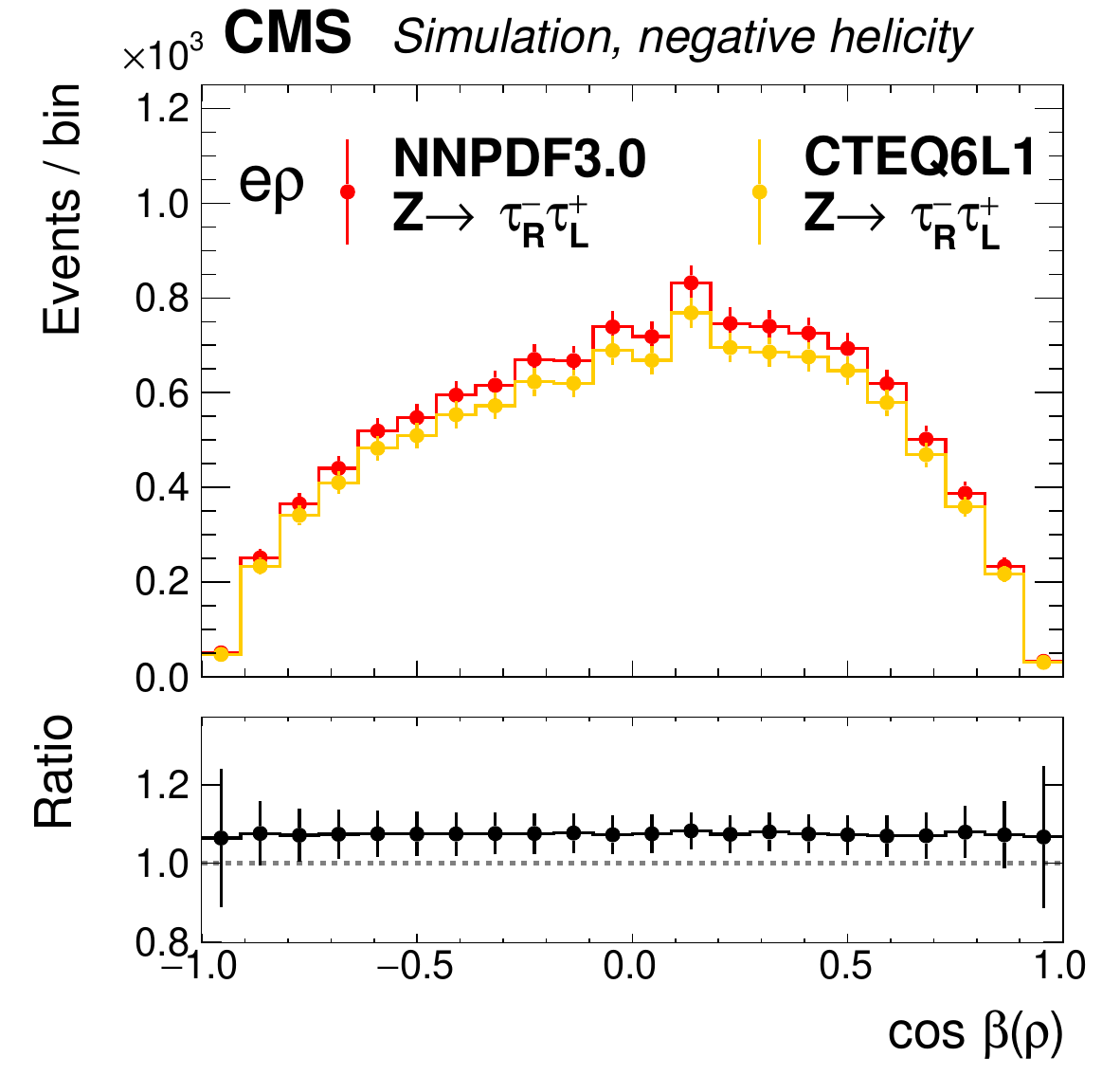}
    \includegraphics[width=0.32\textwidth]{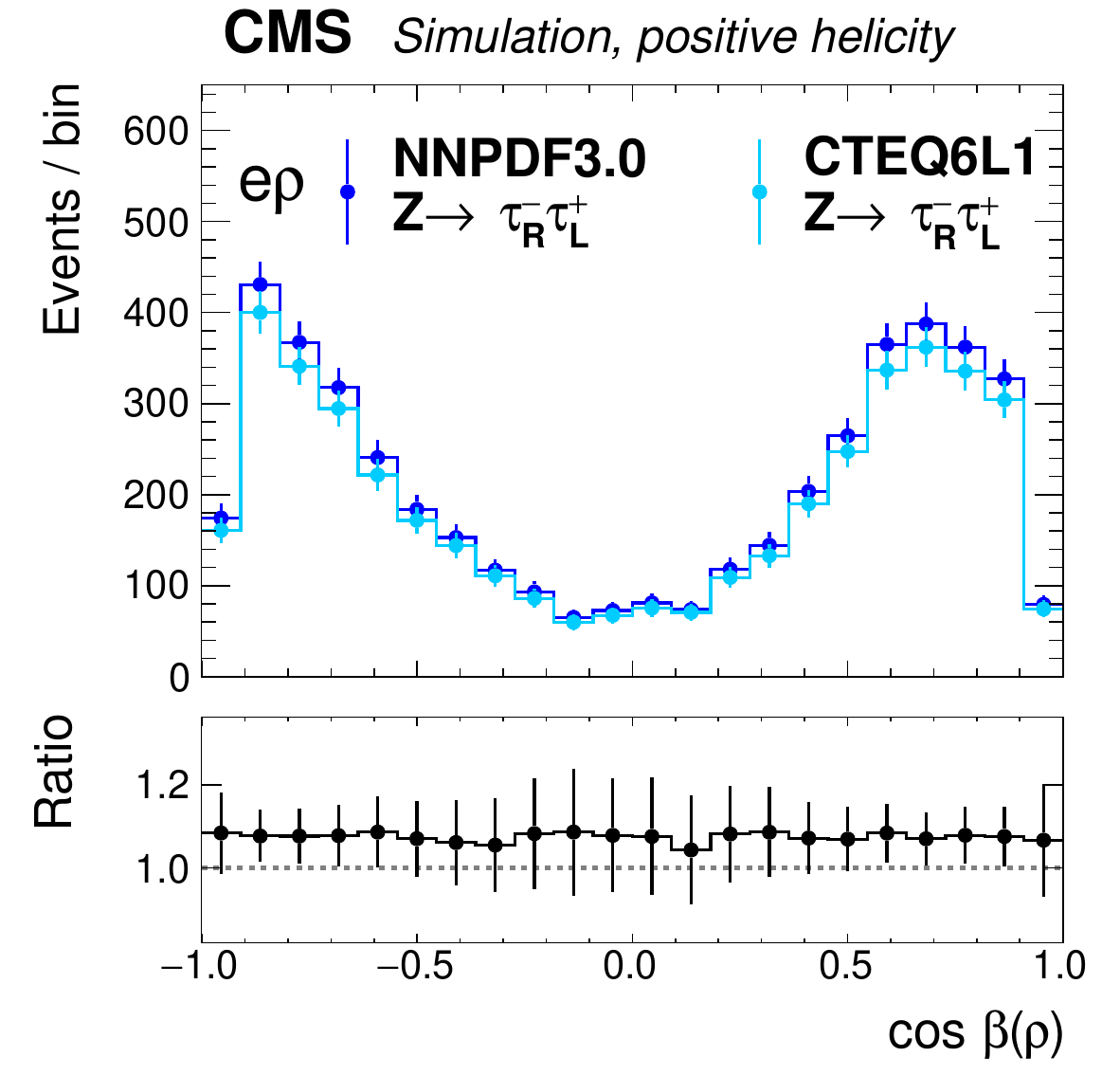}
    \includegraphics[width=0.32\textwidth]{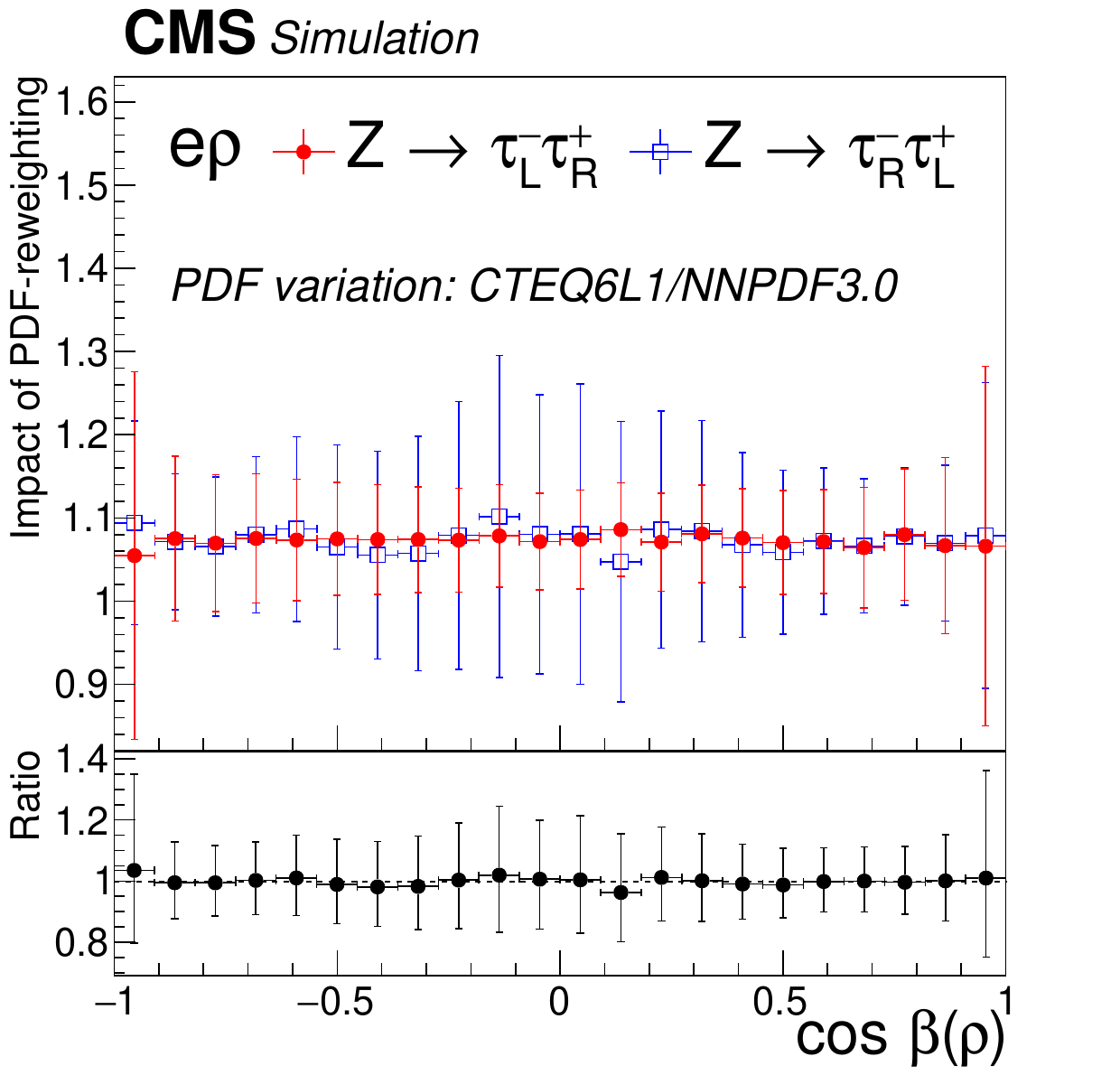}\\
    \includegraphics[width=0.32\textwidth]{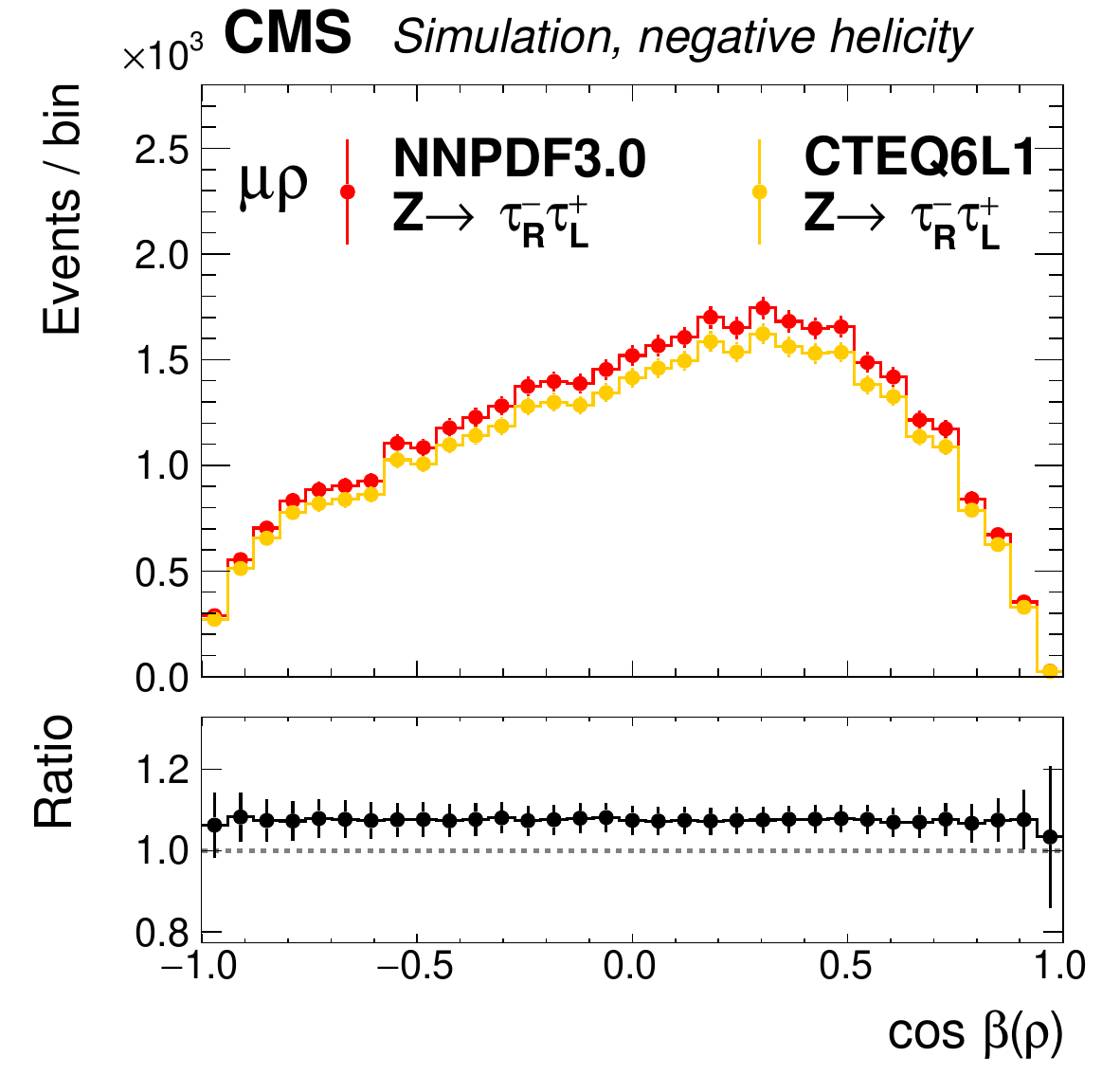}
    \includegraphics[width=0.32\textwidth]{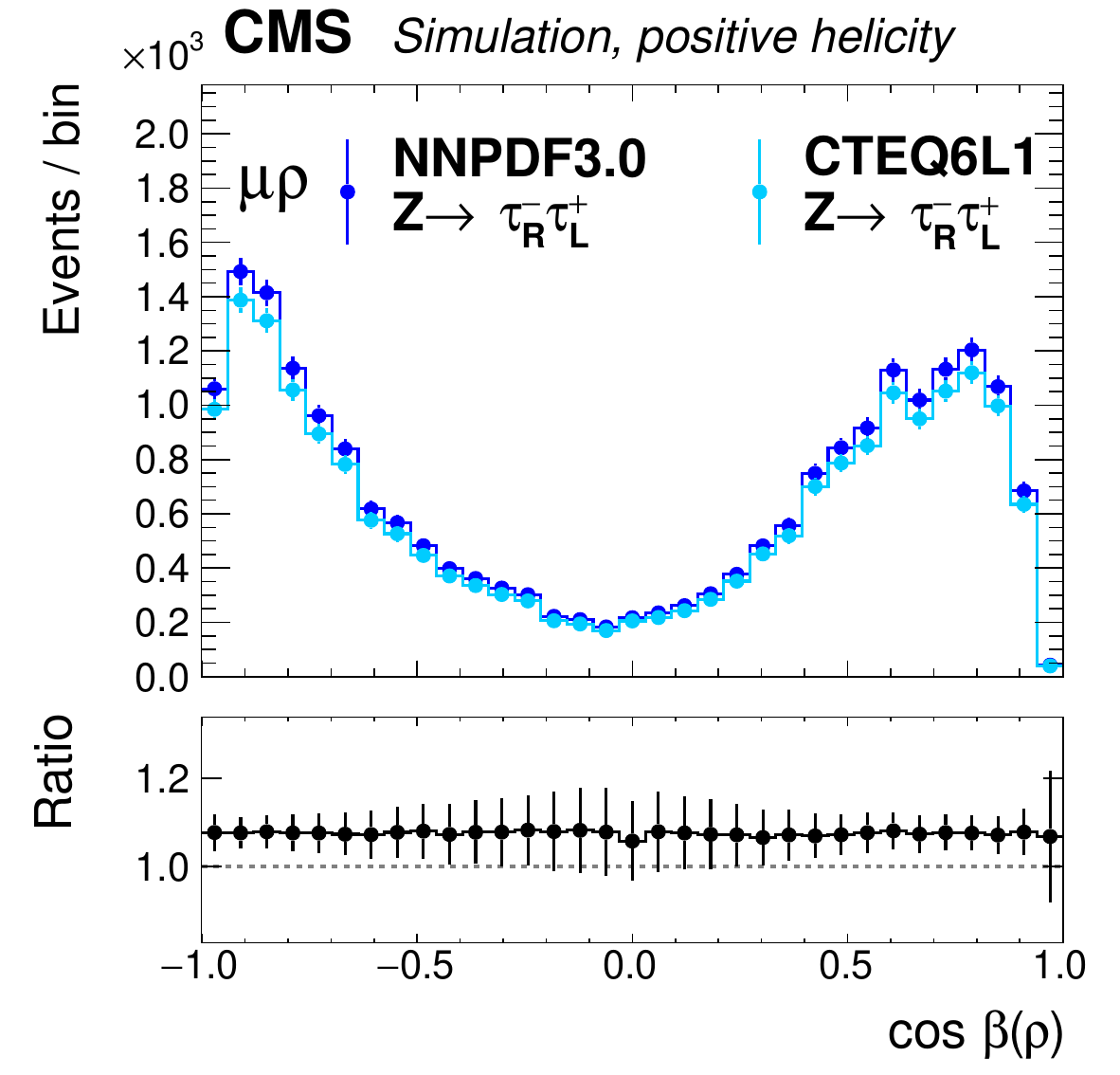}
    \includegraphics[width=0.32\textwidth]{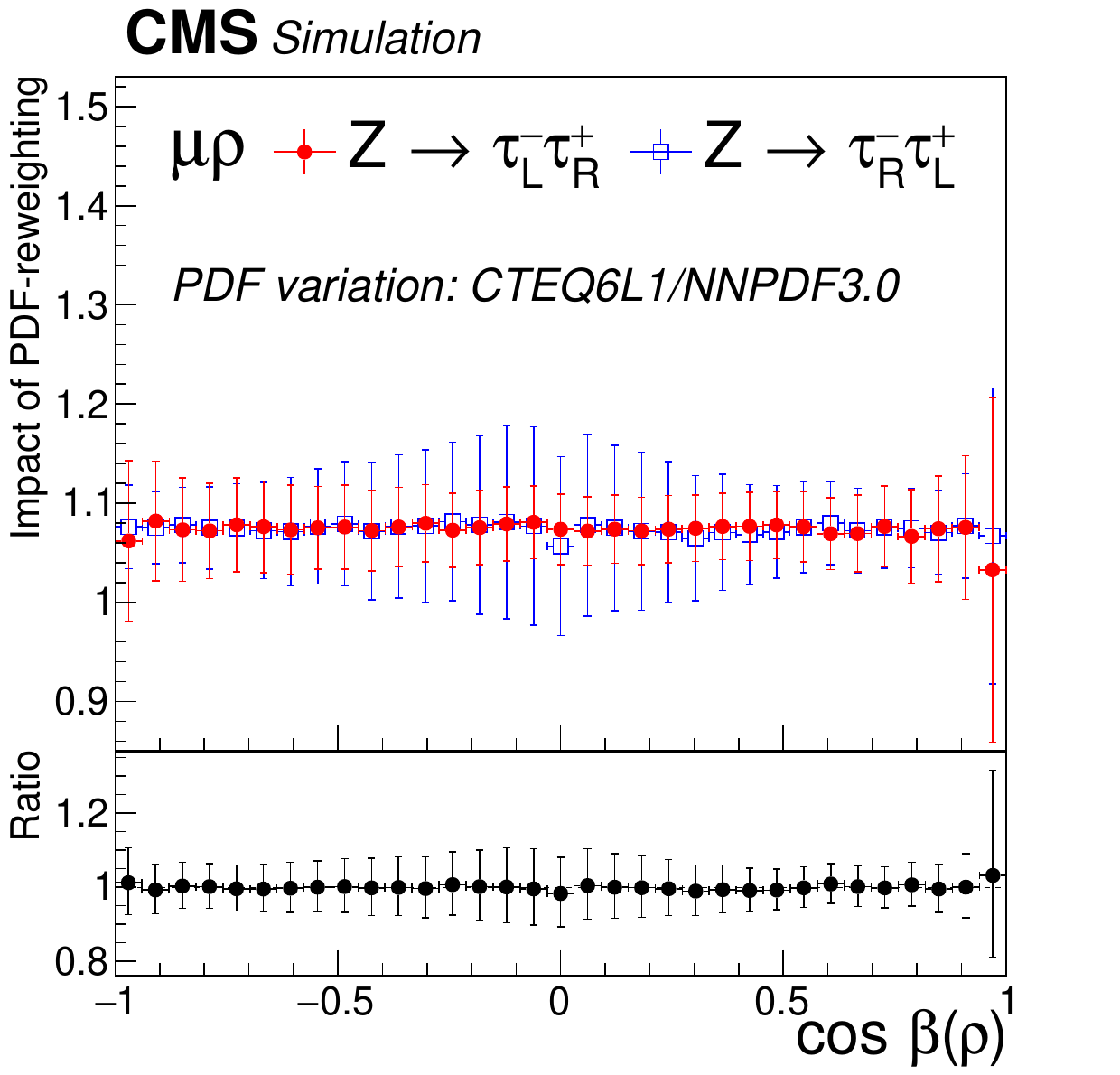}
\caption{Variation of the templates for negative (left) and positive (middle) \tauleptonminus helicities in the $\Pe+\PGr$ (upper row) and $\PGm+\PGr$ (lower row) channels after a change of the PDFs from NNPDF3.0 to CTEQ6L1. The graphs on the right show the ratio of the ratios of the changes for negative and positive helicities, which is flat and centered exactly at 1 demonstrating that a possible PDF effect on the polarization is very small. The drawn error bars are large because the events in the ratios are correlated, so only the fluctuations of the points are relevant and are within approximately 1\%, much smaller than other systematic uncertainties in this analysis.}
\label{temp_change_pdf}
\end{figure}

The efficiencies for right- and left-handed \tauleptonsminus are different and depend strongly on the decay mode and the selection applied in the analysis, especially on transverse momenta of the \tauleptonminus decay products.
Therefore, we adopt a method where these effects are included by using templates for right- and left-handed \tauleptonsminus. The templates are histograms of the $\PGtm$ polarization observables for each decay category of the \tauleptonminus.  They are generated by applying all selections and corrections of the analysis chain to the simulated events of the signal. 
Generator-level information is used to construct
separate templates for right- and left-handed \tauleptonsminus. 

These templates are normalized such that they correspond to zero polarization for a given range (75--120\GeV) of $\sqrt{\hat{s}}$ at the generator level. 
The normalization proceeds in the following way. 
First the templates are defined using the simulated data samples described above for signal and background:
\begin{linenomath*}\begin{equation}
\label{equation_signal_plus_background_is_data}
\mathcal T(\mathrm{simulation}) = \mathcal T(\text{sig.}, \averagepolarisation, r) + \mathcal T(\text{bkg.}).
\end{equation}\end{linenomath*}
The signal template $\mathcal T(\mathrm{sig.}, \averagepolarisation, r)$  contains the two parts $\mathcal T(\ztautaupospol)$ 
and $\mathcal T(\ztautaunegpol)$ for the two helicity states and depends on two parameters of interest, the average \PGt polarization \averagepolarisation and an overall signal strength modifier $r$ as given in the following equation,
\begin{linenomath*}\begin{equation}
\label{equation_signal_parametrisation}
\mathcal T(\mathrm{sig.}, \averagepolarisation, r) = r \left[ \frac{1+\averagepolarisation}{2} \mathcal T(\ztautaupospol) + \frac{1-\averagepolarisation}{2} \mathcal T(\ztautaunegpol) \right].
\end{equation}\end{linenomath*}
To extract the average \tauleptonminus polarization directly from the fit, the signal templates entering the fit must be rescaled such that they correspond to an unpolarized sample of \ztautau events. 
This is achieved by renormalizing the integrals of the templates $\mathcal T(\ztautauposnegpol)$ with the product of 
two factors, $s_{\mathrm{LR}/\mathrm{RL}}$ and $s_\mathrm{tot}$, where
\begin{align} 
s_{\mathrm{LR}/\mathrm{RL}} &= \frac{1}{N_\text{gen}(\ztautauposnegpol)} \label{equation_unpolarisation_factors}, \\ 
s_\mathrm{tot} &= N_\text{gen}(\ztautaupospol) + N_\text{gen}(\ztautaunegpol) \label{equation_integral_factor}.
\end{align}
These normalization factors are obtained from event counts at the generator level for each $N_\mathrm{gen}$ before any selections are applied. The scale factor  $s_{\mathrm{LR}/\mathrm{RL}}$ produces an unpolarized sample before the event selection and the second factor $s_\mathrm{tot}$ ensures that the overall number of signal events remains unchanged. 
In this way, the templates for right- and left-handed helicities incorporate all
biases due to the trigger and analysis selections, and 
the average polarization can be read off directly from the fit results according to Eq.~(\ref{equation_signal_parametrisation}).

All experimental selections are contained in the templates and all systematic uncertainties discussed in the previous section are applied simultaneously to the signal and background templates in the global maximum likelihood fits presented in Section~\ref{results}. The significance of the specific normalization of the templates means that they refer to a null average polarisation within a mass window of 75 to 120\GeV at the generator level.

A global maximum-likelihood fit is performed in the 11 event categories given in Table~\ref{table_disciminators_per_event_categories} in 
the \tauel\taumu, \tauel\tauhad, \taumu\tauhad, and \tauhad\tauhad channels. We treat the signal strength $r$ as a free parameter in this fit in order to ensure that only the shape information is used to estimate the average \taulepton polarization \averagepolarisation. 

The templates of the two signal contributions \ztautauposnegpol reflect the bias from detector acceptance and event selection, and thus the average polarization is retrieved correctly by the fit according to Eqs.~(~\ref{equation_signal_plus_background_is_data}) and~(~\ref{equation_signal_parametrisation}).

A closure test using MC simulated events was performed to verify that the procedure for measuring the \PGt lepton polarization is unbiased.

Representative templates for three of the 11 decay categories are presented in Fig.~\ref{fig_template}. 
The difference in the templates between right- and left-handed \tauleptonsminus is clearly visible. 
The distinctive shapes in different decay channels are partially distorted by the trigger and analysis selection. This is particularly true for the lepton plus pion channel where the $\pt$ selection suppresses the left-hand part of the distribution. From this, it is evident that various categories have different discrimination power.   
The distributions also demonstrate some bin-by-bin fluctuations that sometimes exceed the statistical fluctuations of the MC samples. Those fluctuations are included when treating the systematic uncertainties, whereas the plots contain only the statistical uncertainties.    

To test the stability of the templates with respect to different PDF choices, events are reweighted by the relative weights of 13 different sets of PDFs available for the analysis and new templates are created from the reweighted events.
The change in the normalization and shape of the templates was studied with respect to the nominal PDF set (NNPDF3.0), finding that 
only the overall normalization changes by at most a few percent, whereas the shapes remain unchanged, affecting the templates for positive and negative helicity in the same way, so that the relative ratio between the templates remained constant. 
In other words, the average polarization extracted from the ratio of templates according to Eq.~(\ref{equation_signal_parametrisation}) does not change. 

In Fig.~\ref{temp_change_pdf} we compare templates with NNPDF3.0 and CTEQ6L1 PDFs for both negative and positive \tauleptonminus helicities in the $\Pe+\PGr$ and $\PGm+\PGr$ categories.
The change is uniform and of a few percent, and 
the double ratio of the relative overall changes is identical to unity for all studied variations and less than 1\% on the polarization measurement, as shown by the third graph of Fig.~\ref{temp_change_pdf}. 

A further test of the stability of the procedure consisted in a subdivision of the data into three bins of 
pseudorapidity of the decaying \PZ boson: $\abs\eta^{\PZ}<1.3$, $1.3<\abs{\eta^{\PZ}}<2.2$, and $\abs{\eta^{\PZ}}>2.2$. This test was of particular importance to assure that 
there is no bias in the polarization measurement in different kinematic regions.

\section{Results}
\label{results}
\begin{figure}[htbp]
     \centering
     \includegraphics[width=0.47\textwidth]{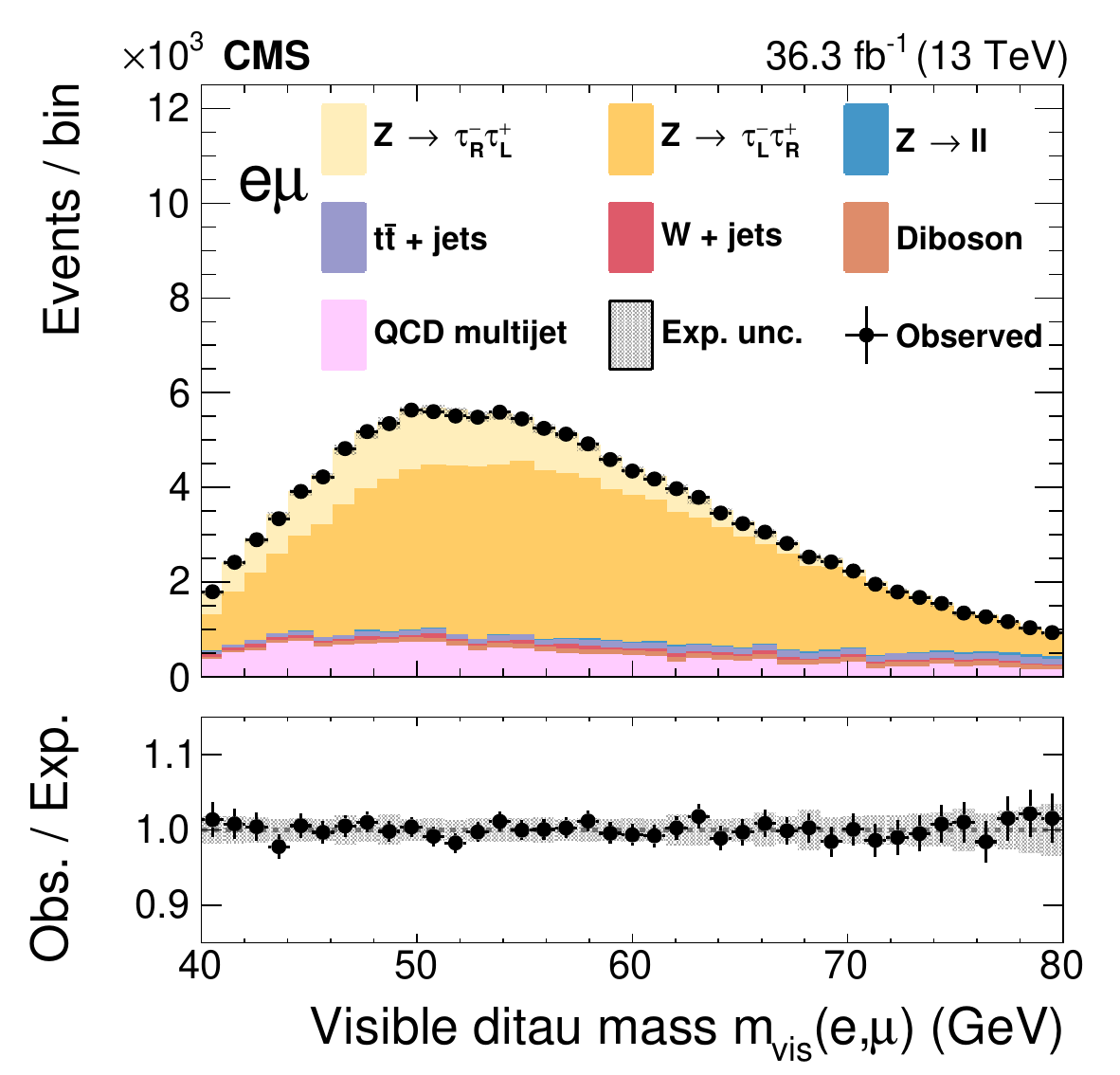}
   \caption{The final maximum likelihood fit of the template to the data for the $\elmu $ channel. The figure shows the contributions of all backgrounds and the fitted contributions from \tauleftminus and \taurightminus helicity states. The bottom panel of shows the ratio of the experimental data to the sum of the expected contributions. The vertical error bars on the data points are the statistical uncertainties, the gray rectangles indicate the systematic experimental uncertainty in the total expectation. }
   \label{fig_elmu}
  \end{figure}

\begin{figure}[htbp]
     \centering
     \includegraphics[width=0.47\textwidth]{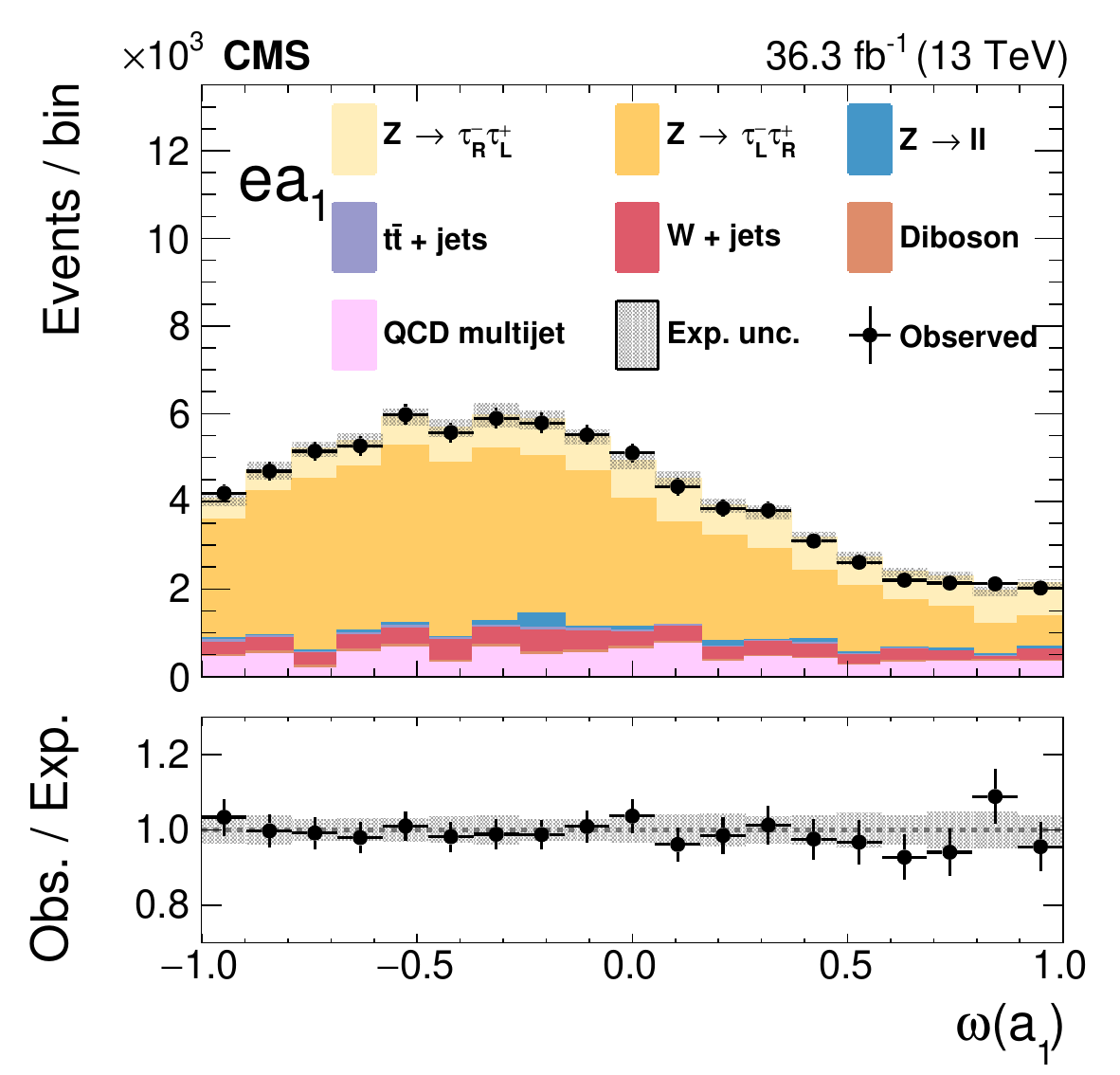}
     \includegraphics[width=0.47\textwidth]{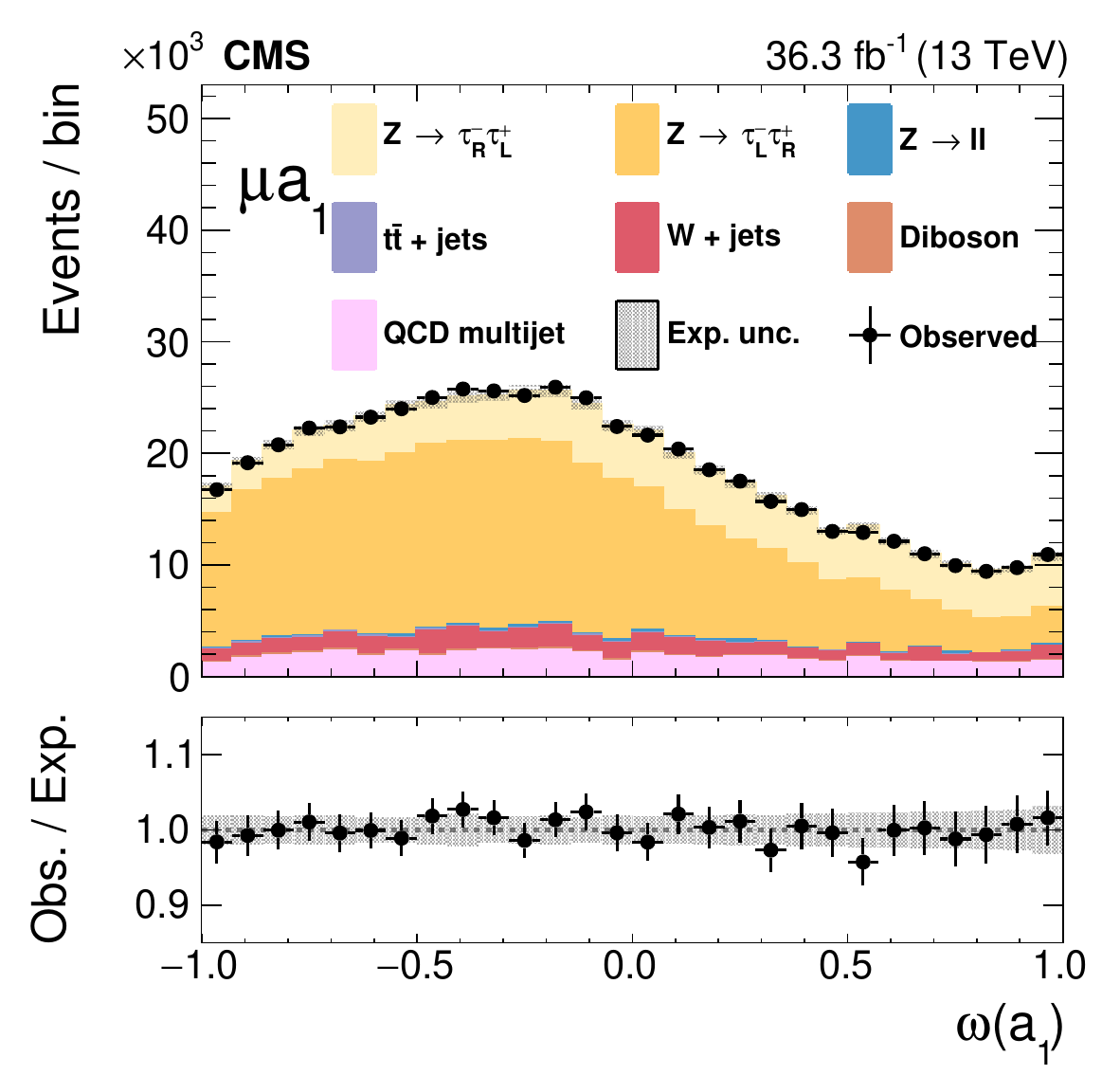} \\
     \includegraphics[width=0.47\textwidth]{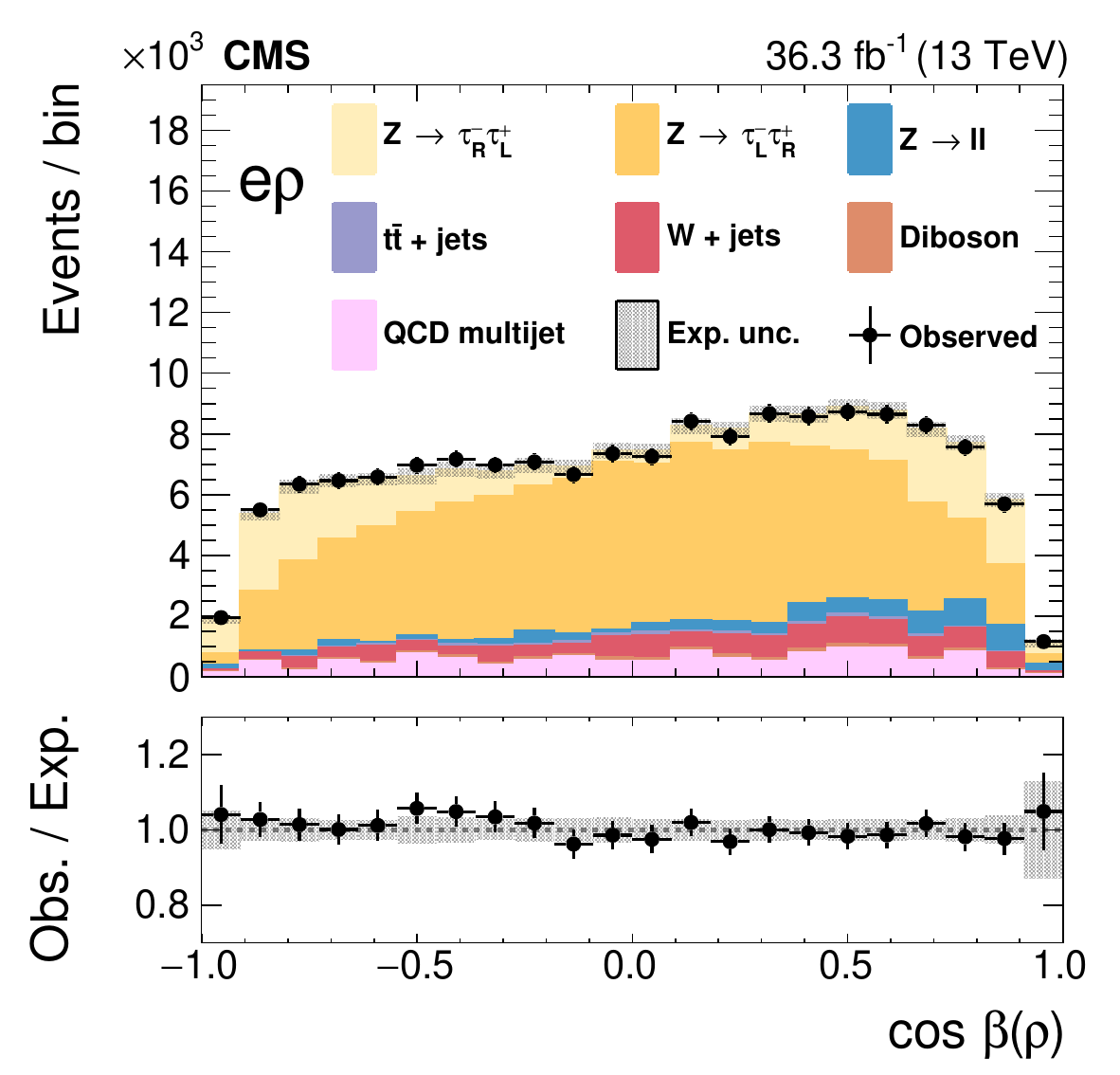}
     \includegraphics[width=0.47\textwidth]{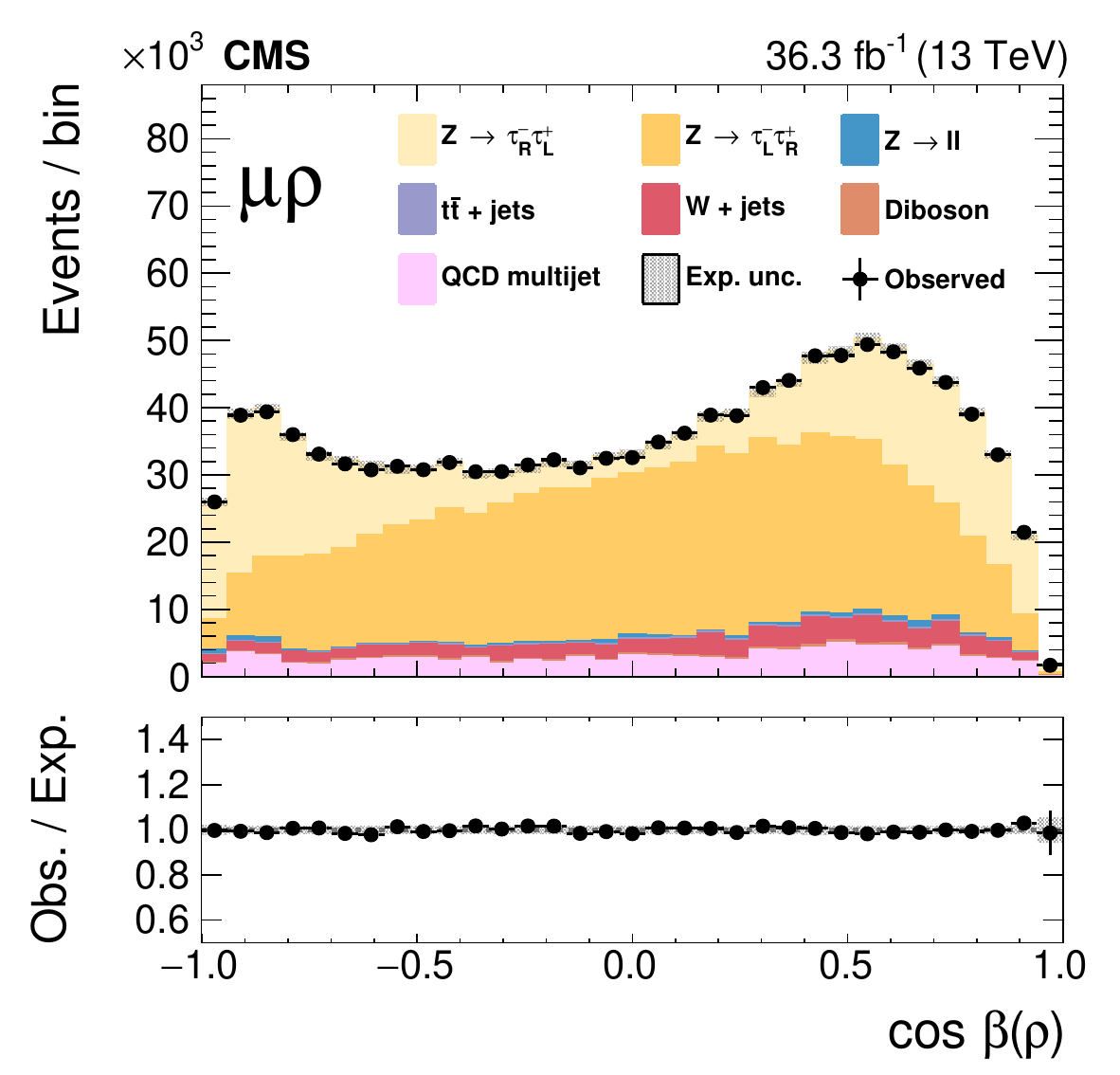} \\
     \includegraphics[width=0.47\textwidth]{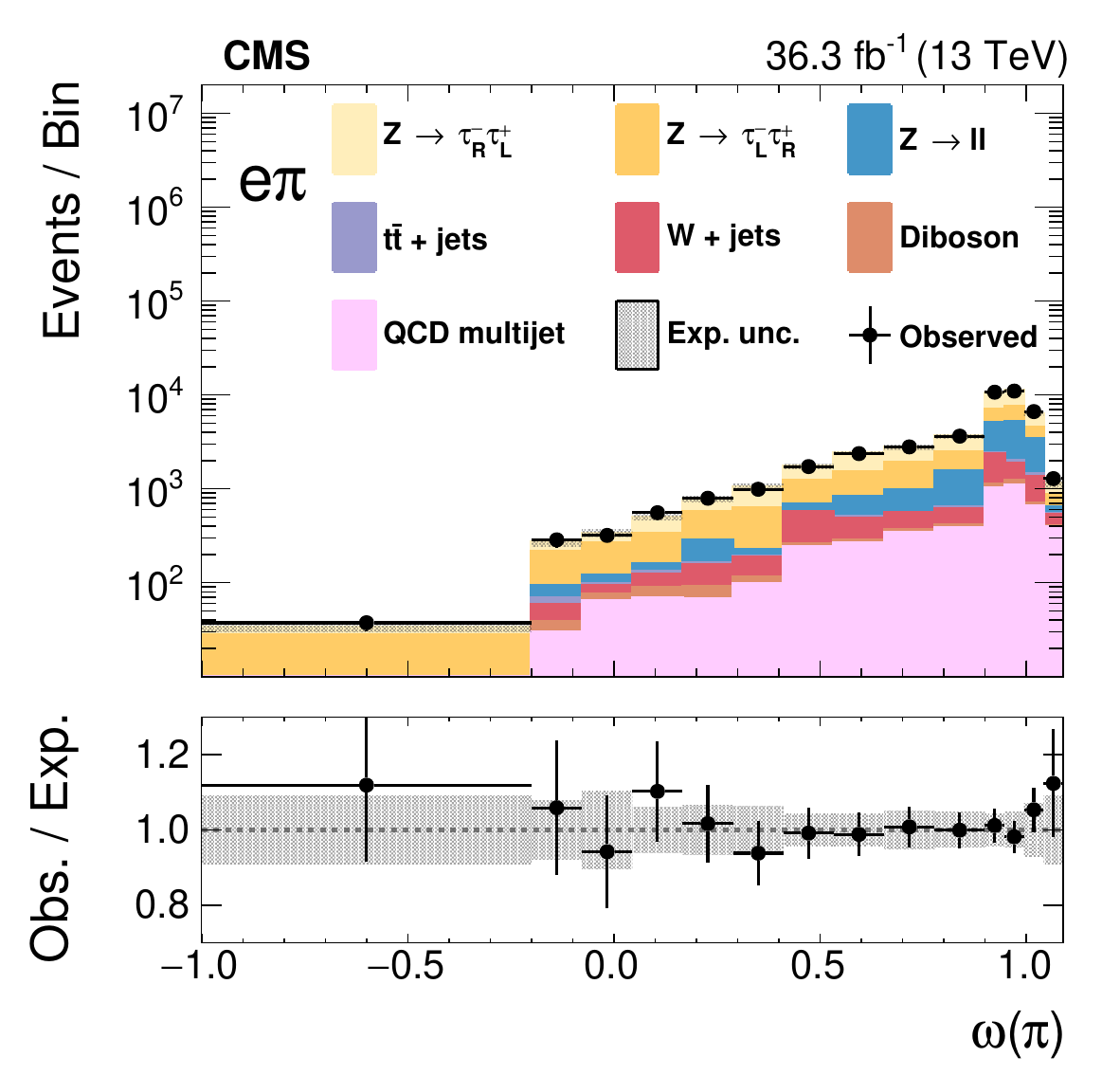}
     \includegraphics[width=0.47\textwidth]{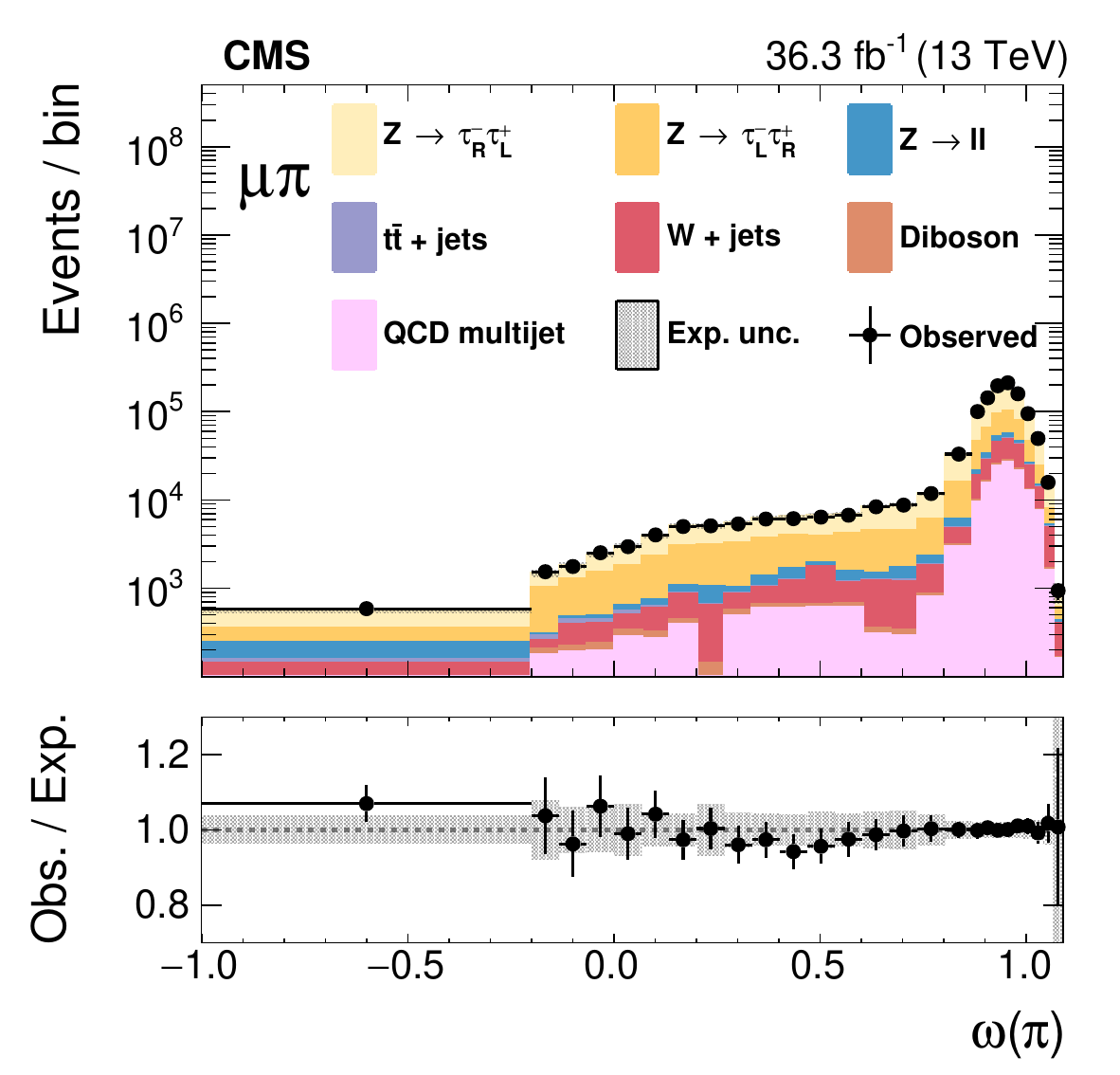}

     \caption{The final maximum likelihood fits of templates to the data for the 
   three $\elhad$ (left) and three $\muhad$ (right) categories. The figures show the contributions of all backgrounds and the fitted contributions from \tauleftminus and \taurightminus helicity states. The bottom panels show the ratio of the experimental data to the sum of the expected contributions. The vertical error bars on the data points are the statistical uncertainties, the gray rectangles indicate the systematic experimental uncertainty in the total expectation.}
   \label{fig_elhad}
 \end{figure}

\begin{figure}[htb]
    \centering

     \includegraphics[width=0.47\textwidth]{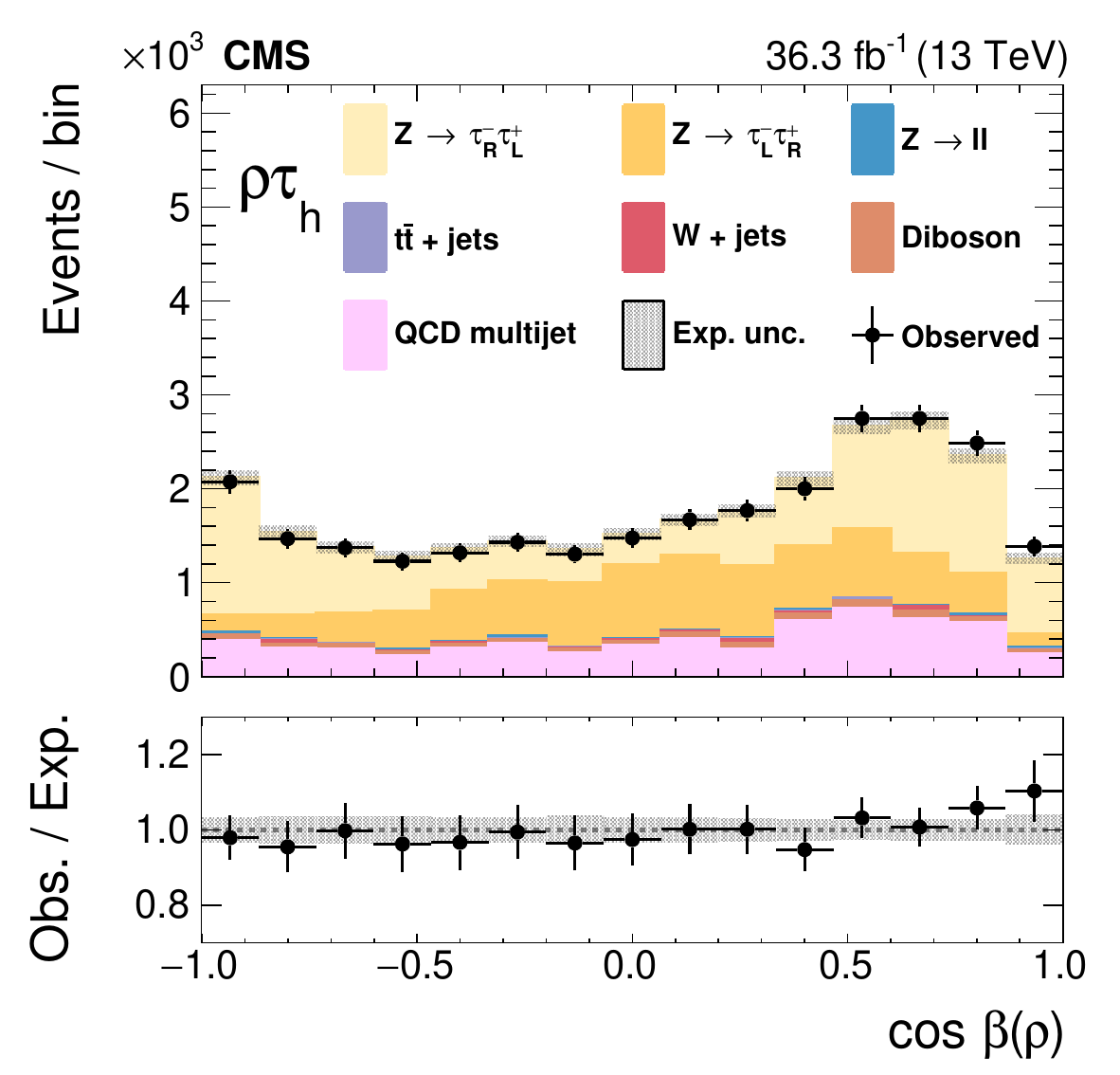}
     \includegraphics[width=0.47\textwidth]{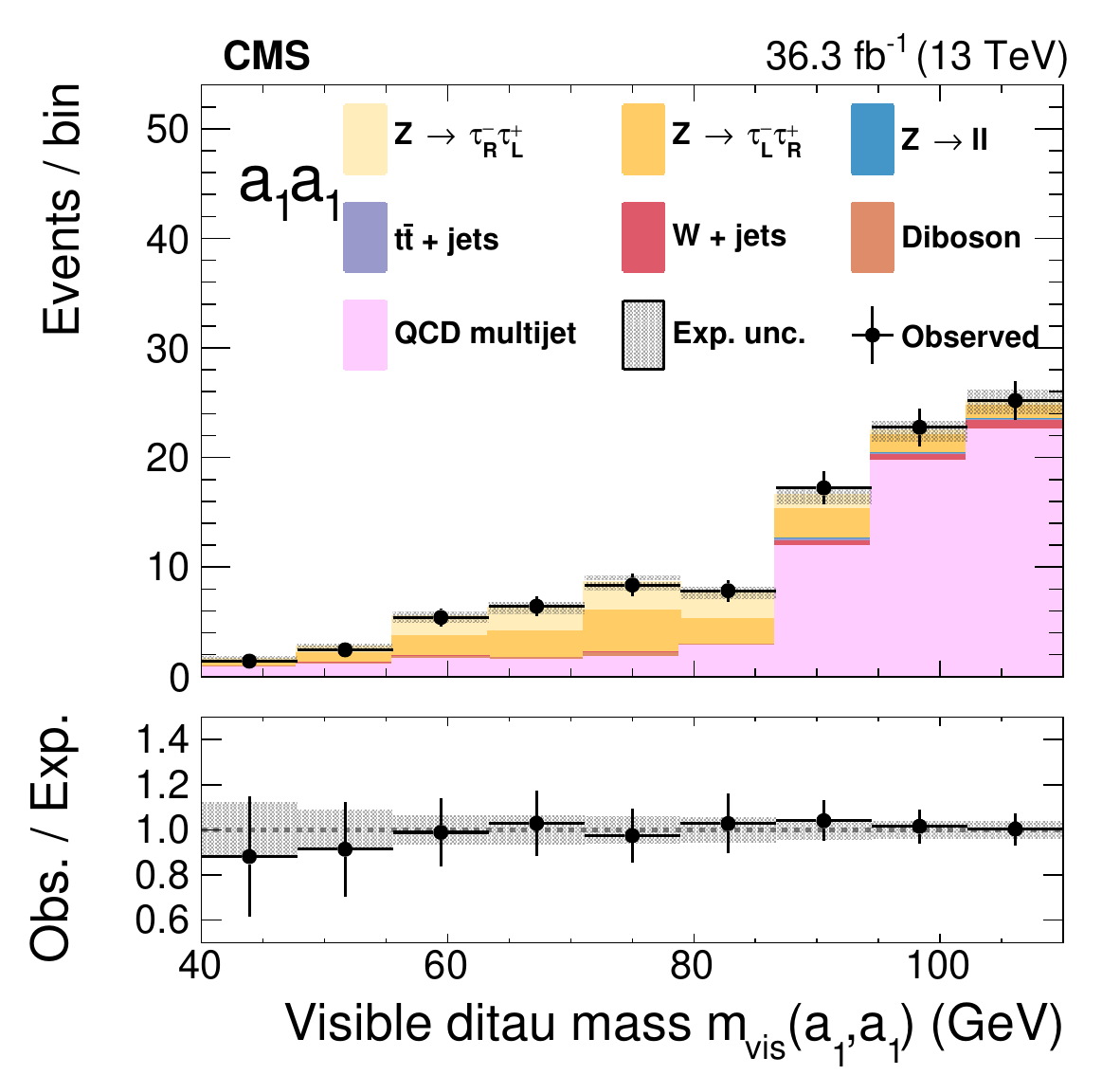} \\
     \includegraphics[width=0.47\textwidth]{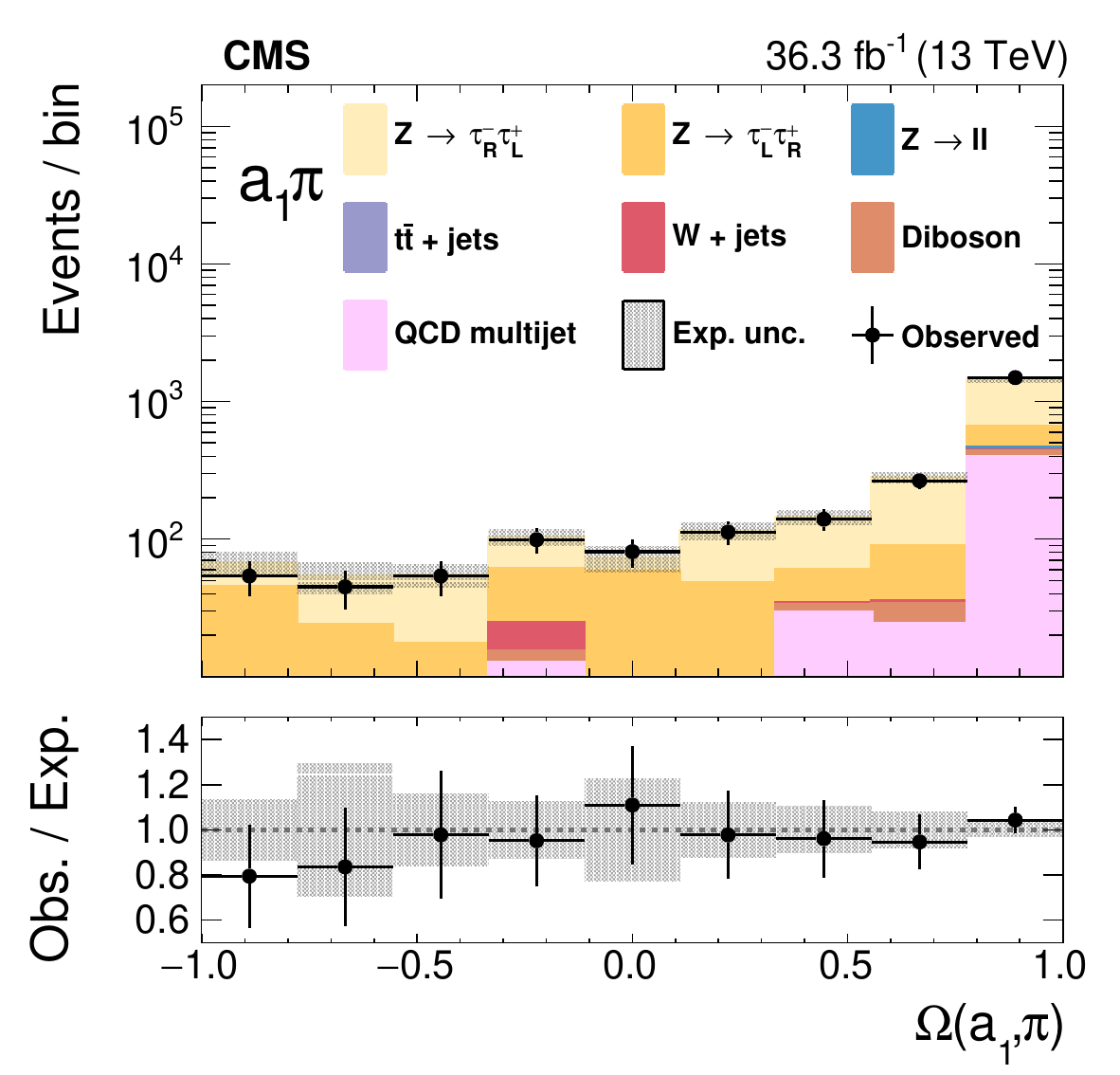}
     \includegraphics[width=0.47\textwidth]{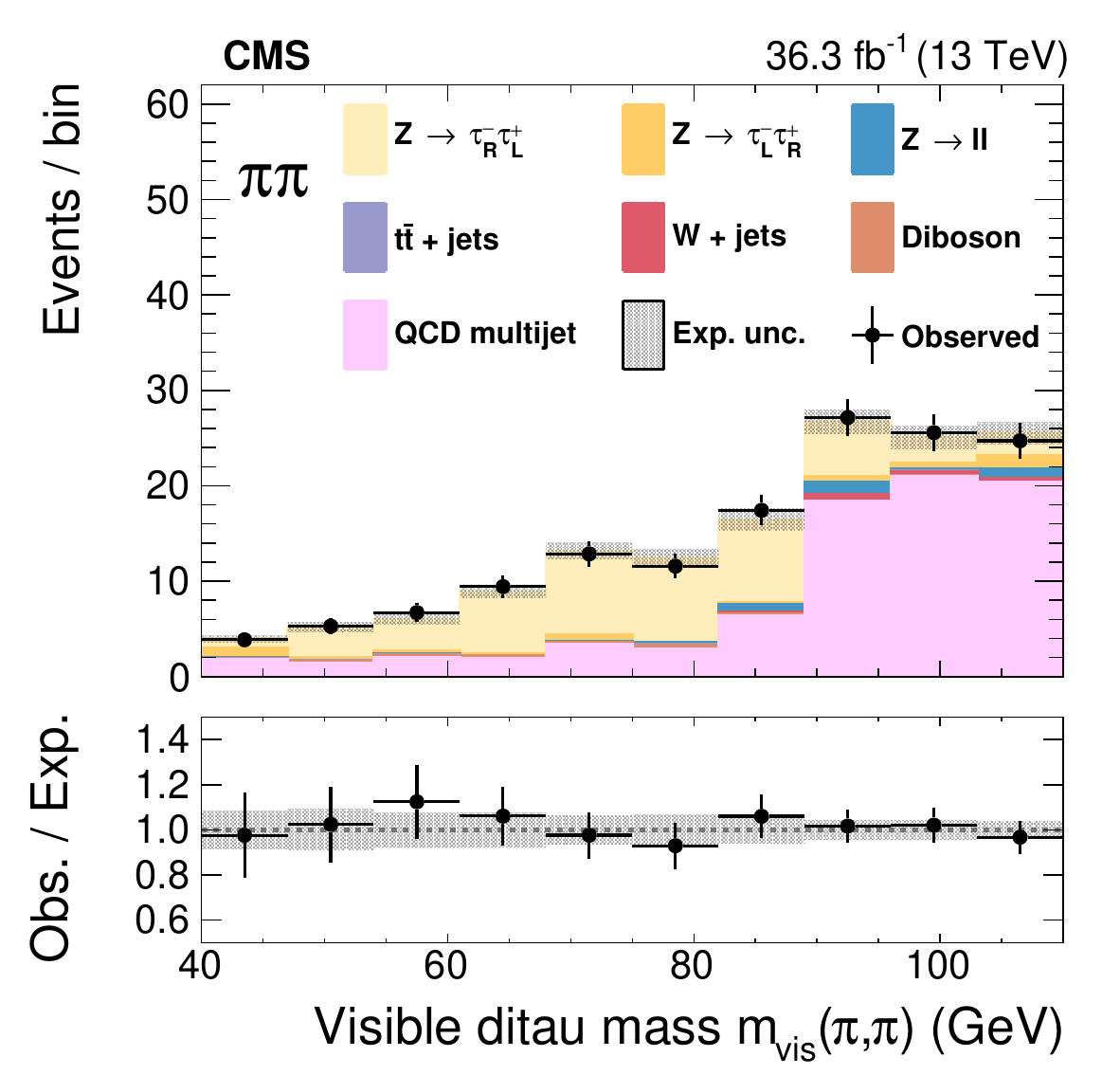}
   \caption{The final maximum likelihood fits of the templates to the data for the $\hadhad$ channels. The figures show the contributions of all backgrounds and the fitted contributions from \tauleftminus and \taurightminus helicity states. The bottom panels show the ratio of the experimental data to the sum of the expected contributions. The vertical error bars on the data points are the statistical uncertainties, the gray rectangles indicate the systematic experimental uncertainty in the total expectation.}
   \label{fig_hadhad}
 \end{figure}

In this section we present the results of the maximum likelihood fits of the templates to the observed distributions of the chosen discriminators (see Table~\ref{table_disciminators_per_event_categories}), which lead to the extraction of the average polarization.
Figures~\ref{fig_elmu},~\ref{fig_elhad}, and~\ref{fig_hadhad}  show how well the data are described by the final maximum likelihood fits
for the categories  \elmu, \elhad and \muhad combined, and \hadhad, respectively. 

 
\begin{figure}[htb]
 \centering   
  \includegraphics[width=0.49\textwidth]{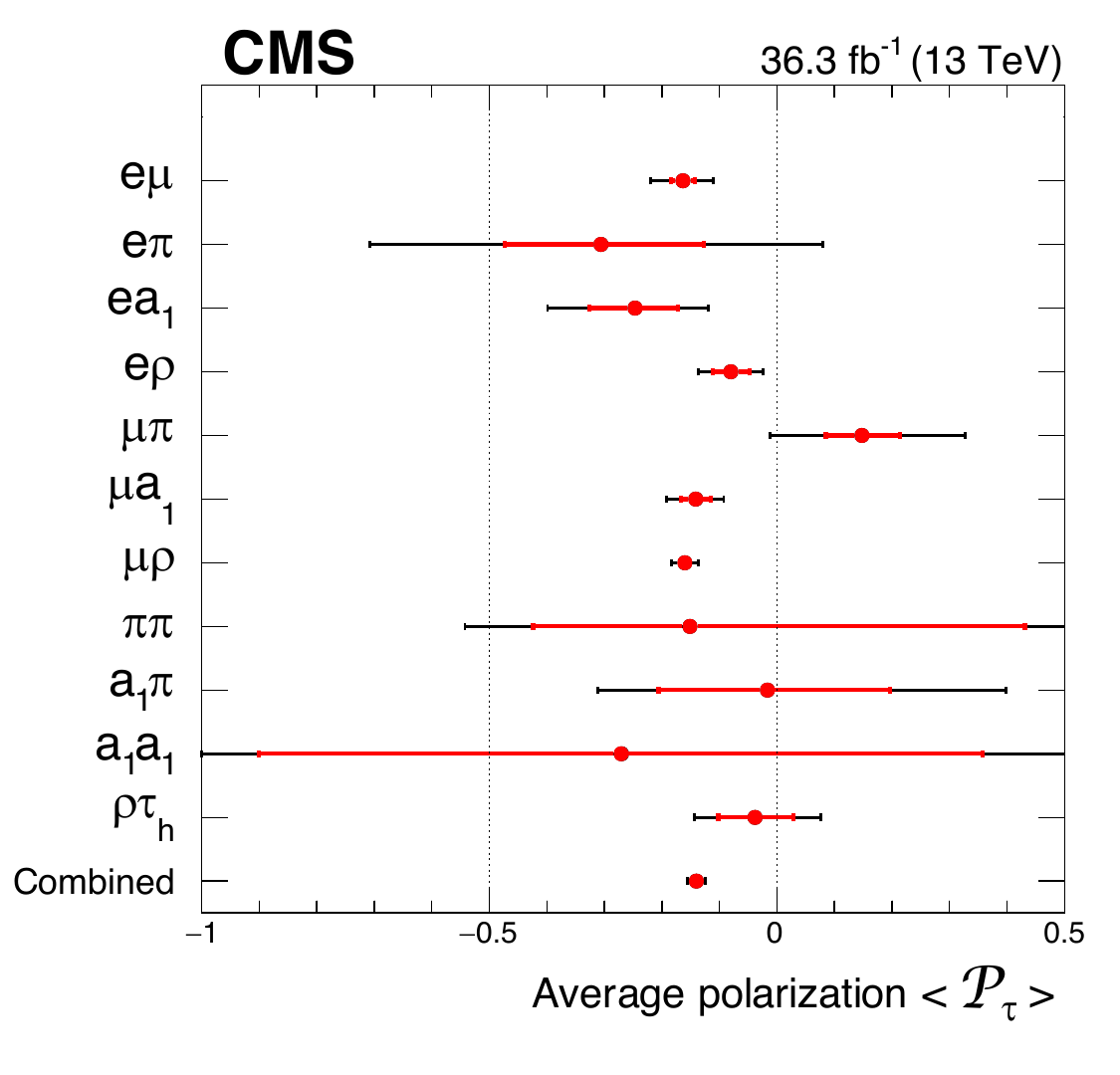}
  \hfill
    \includegraphics[width=0.49\textwidth]{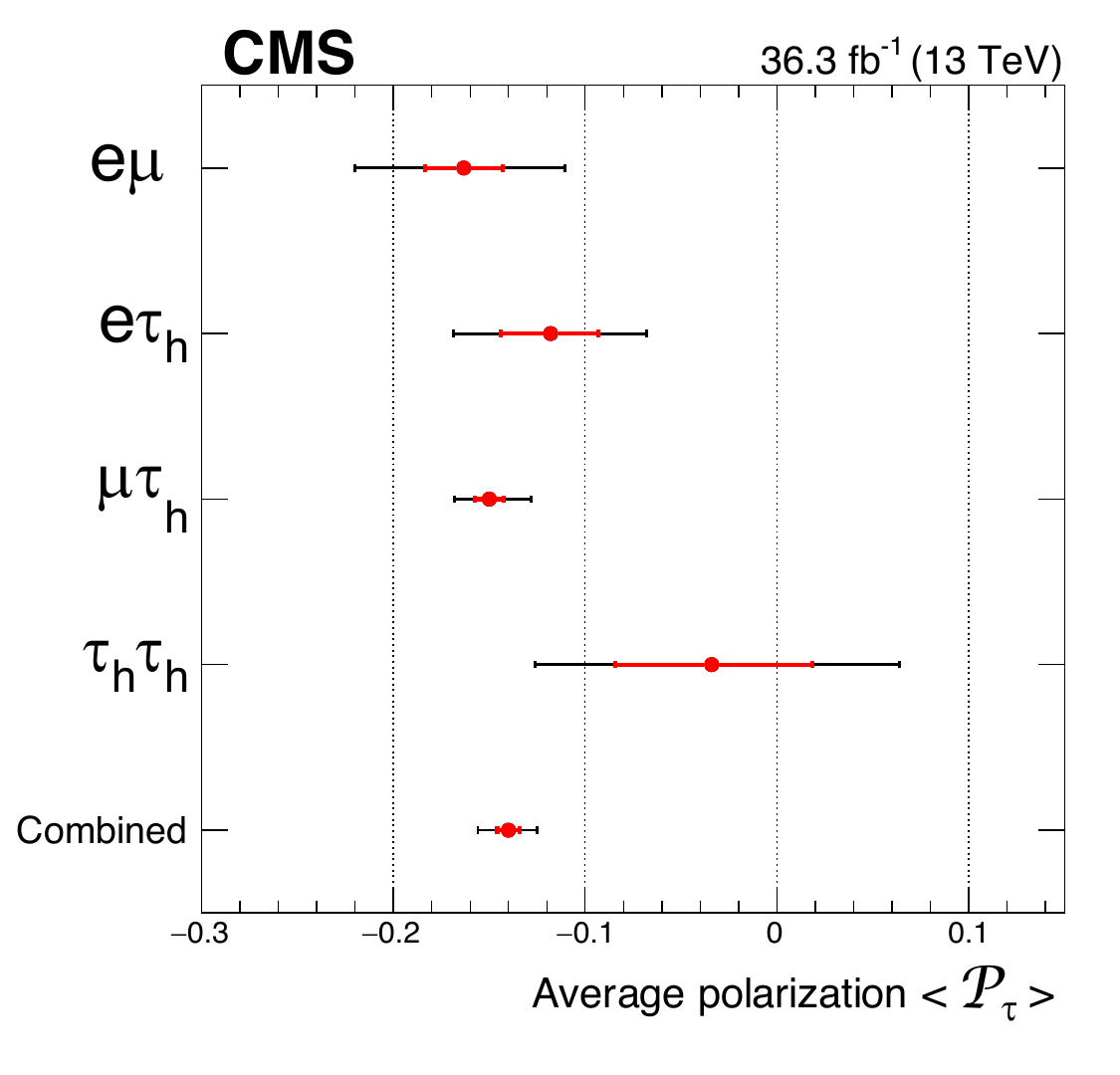}
  \caption{Results of the maximum likelihood fits for the average \tauleptonminus polarization for the 11 event categories and the combined fit as the lowest point in the figure on the left. 
  On the right the results for the categories are grouped into 4 channels separately and are shown together with the combined fit. The inner error bars represent the statistical uncertainty, and the outer bars include the systematic uncertainty.}
 \label{final_pol_data}
\end{figure}

The final results of the average polarization $\langle \mathcal{P}_{\PGt} \rangle_\text{75--120\GeV}$ obtained by these maximum likelihood fits to data are shown in Fig.~\ref{final_pol_data}. 
The subscript 75--120\GeV refers to the normalization of the templates as explained in Section~\ref{section_polarisation_measurement}.
The highest sensitivity to $\mathcal{P}_{\PGt}$ occurs in the \mtCombinedRhoOneprong category, which benefits from more data and 
an optimal observable $\omega_\text{vis}$ based only on directly measured quantities; the lowest sensitivity occurs in the fully hadronic decay categories, which suffer from high trigger and selection thresholds. 

The average \tauleptonminus polarization referring to an interval of 75--120\GeV of the generated di-$\PGt$ mass is:
\begin{linenomath*}\begin{equation}
\langle \mathcal{P}_{\PGt} \rangle_{75-120\GeV} 
= -0.140 \pm 0.006\, \text{(stat)}\, \pm 0.014\syst = -0.140 \pm 0.015.
\end{equation}\end{linenomath*}

The extracted polarization value is stable with respect to a change of the kinematical region like pseudorapidity. 
The polarization is evaluated
in three different $|\eta^{\PZ}|$ bins of the \PZ boson and the values agree with each other within their uncertainties.

The measurement is limited by systematic uncertainties, predominantly the decay mode migrations that contribute an uncertainty of $\pm$0.008 in the global maximum likelihood fit to the average polarisation. An additional important contribution of $\pm$0.010 originates from bin-to-bin fluctuations in the signal MC-data. Other sources of uncertainty contribute less than $\pm$0.005 to the average polarisation.

To compare to previous measurements, such as those at LEP we have to correct the measured average
polarization \pol of the mass interval of 75--120\GeV to the polarization value at the \PZ pole. 

\begin{figure}[hbt] 
\centering
\includegraphics[width=0.7\textwidth]{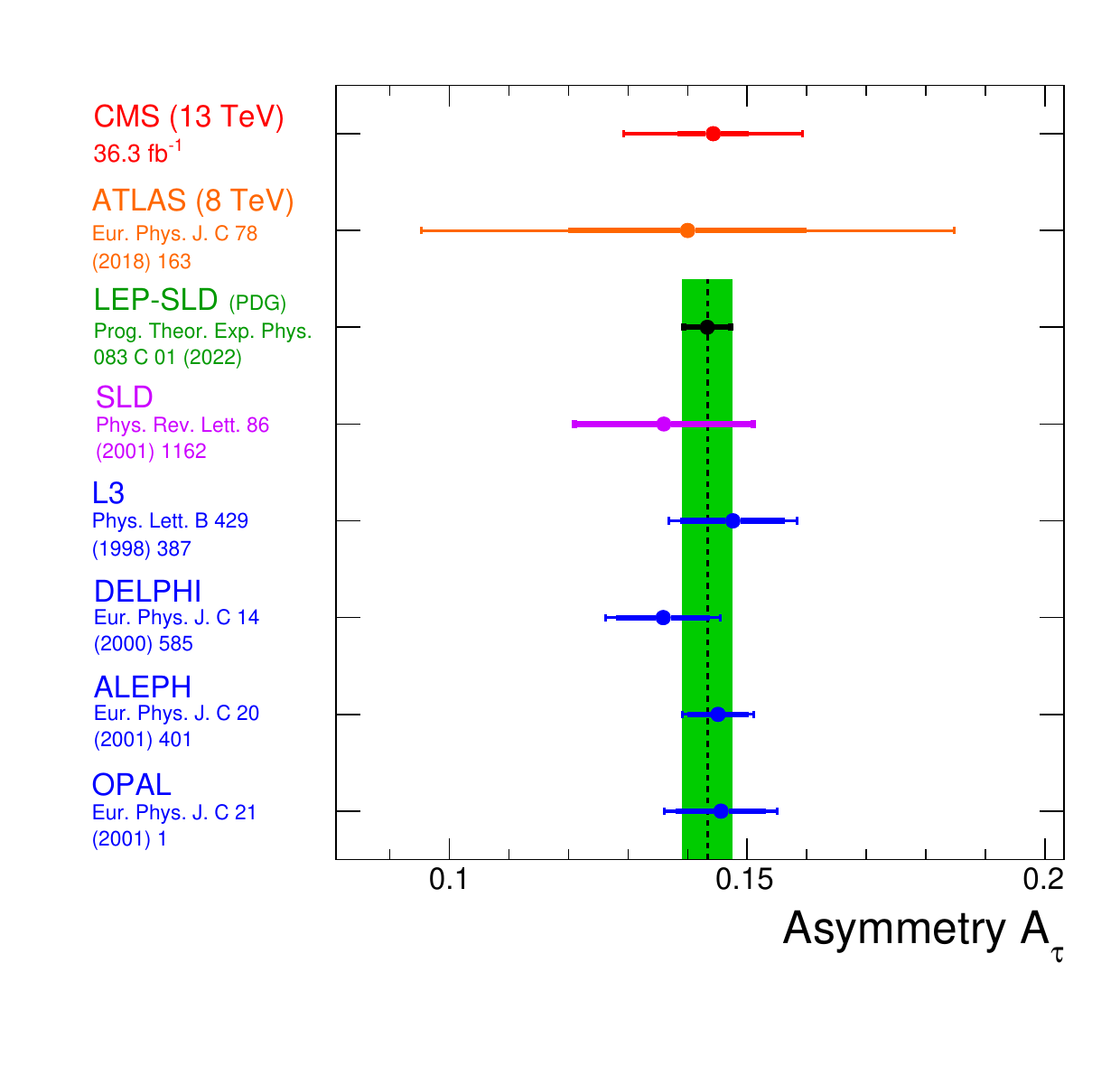}
\caption{A comparison of the  \tauleptonminus asymmetry, $A_{\PGt}$ measured from the \tauleptonminus polarization in this paper and other measurements. The value of $A_{\PGt}$ for CMS is obtained based on the \PZ boson polarization Eq.~(\ref{taupolresult}) and using Eq.~(\ref{consts1}).  The green band 
indicates the \tauleptonminus polarization value obtained by combining the SLD measurement~\cite{PhysRevLett.86.1162} 
with the measurements performed at LEP (ALEPH~\cite{Heister:2001uh}, DELPHI~\cite{Abreu:1999wv}, L3~\cite{Acciarri:1998vg}, and OPAL~\cite{Abbiendi:2001km}).
The measurement performed by the ATLAS Collaboration at a lower center-of-mass energy of 8 TeV is documented in Ref.~\cite{Aaboud:2017lhv}. The CMS measurement refers to the result of the analysis presented in this paper. The inner horizontal bars represent the statistical uncertainly, the outer bars include the systematic uncertainty. }
\label{figure_data_comparision_LEP_SLAC_LHC}
\end{figure}

To establish a relation between the average polarization and the value at the \PZ pole we generated four \MGvATNLO samples of Drell--Yan events at LO in EWK with different values for the weak mixing angle.
For each of these samples of 1 million events, we count the number of left- and right-handed \tauleptonminus within the interval $75\leq\sqrt{\hat{s}}\leq120\GeV$ at the generator level. 
In this way we calculate the average polarization according to its definition,  Eq.~(\ref{equation_definition_polarisation}). We find a constant shift of $ -0.004$ between the average and the peak polarization estimated according to Eq.~(\ref{eqnumber5}), independently of the weak mixing angle.

The corrected value of the polarization at the \PZ pole, which gives directly the negative of the asymmetry parameter $A_{\PGt}$, is then:
\begin{linenomath*}\begin{equation}\label{taupolresult}
\mathcal{P}_{\PGt}(\PZ) = -A_{\PGt}=-0.144 \pm 0.006\stat \pm 0.014\syst = -0.144 \pm 0.015.
\end{equation}\end{linenomath*}

Figure~\ref{figure_data_comparision_LEP_SLAC_LHC} compares the measured asymmetry parameter $A_{\PGt}$ 
with published results of previous experiments: the four LEP experiments~\cite{ALEPH:2005ab},  SLD~\cite{PhysRevLett.86.1162} and the combination of LEP and SLD.
The earlier result of ATLAS at $\sqrt{s}=8\TeV$~\cite{Aaboud:2017lhv}, which is also included in the figure is an average over the mass range of 66--116\GeV and was not corrected to the \PZ pole. 
The LEP experiments used the \textsc{ZFITTER} package~\cite{Akhundov:2013ons} to translate a measured average polarization around the peak of the \PZ resonance to the asymmetry parameter $A_{\PGt}$. 
At LEP the average over a small mass interval was due to initial- and final-state photon radiation. 
Their observed shift of the average polarization with respect to the pole value is accidentally of the same size as the correction applied in this analysis. 
Initial-state gluon radiation of the incoming quarks, which is more important than photon radiation, is included in the \MGvATNLO simulation. 
No further NLO corrections have been applied to the asymmetry parameter, since it is expected that their effect will be significantly smaller than the systematic uncertainties.
It is the most precise measurement of $A_{\PGt}$ at the LHC and of comparable precision as the SLD experiment. 

The measured asymmetry parameter $A_{\PGt}$ by CMS is in good agreement with the LEP average of $A_{\PGt} = 0.1439 \pm 0.0043$~\cite{ALEPH:2005ab} and 
compatible with the global SM value for the lepton asymmetry parameter assuming lepton universality $A_{\ell} = 0.1468 \pm 0.0003$~\cite{Rpp:2022}. 
The tabulated results presented in Figs.~\ref{final_pol_data} and \ref{figure_data_comparision_LEP_SLAC_LHC} are available in HEPData~\cite{hepdata}.

The polarization of \tauleptonsminus produced in \PZ bosons decays is a direct consequence of the structure of the weak interaction, namely the mixing of the $\PBz$ and the $\PW^0$ fields, which is parametrized by the effective mixing angle, \sintwoth. In the analysis it is not possible to distinguish between \ztautau and \gammatautau, because they produce exactly the same final state. The \gammatautau process results in unpolarized \tauleptonsminus. 
Because the relative fraction of \dytautau events changes as a function of the quark-antiquark invariant mass
$\sqrt{\hat{s}}$, the polarization has to be considered as a function of this mass, the effective mixing angle, and the quark type, specifying whether the \PZ has been produced via a pair of up- or down-type quarks, see Eq.~(\ref{pol_i}). This average has been accounted for in our procedure described in Section~\ref{section_polarisation_measurement} to correct the measured polarization to the \PZ pole. 
This result can be directly used in the relation of Eq.~(\ref{eqnumber5}) to obtain a value for the effective electroweak mixing angle:

\begin{linenomath*}\begin{equation}
 \sintwoth = 0.2319 \pm 0.0008\stat\pm 0.0018\syst = 0.2319 \pm 0.0019.
\end{equation}\end{linenomath*}

Our result is in good agreement with the most precise value of the effective weak mixing angle measured in \zll decays obtained by the combination of the LEP and SLD results $\sintwoth = 0.2315 \pm 0.0002$~\cite{ALEPH:2005ab}.

\section{Summary}
\label{sec_sum}
The CMS detector was used to measure the polarization of \tauleptonsminus in the decay of \PZ bosons produced in proton-proton collisions 
at the LHC at $\sqrt{s}=13\TeV$ in a data sample corresponding to an integrated luminosity of 36.3\fbinv. 
Eleven different combinations of decay modes of the \tauleptons were used to study the polarization.

The measured \tauleptonminus polarization, $\mathcal{P}_{\PGt}(\PZ) = -0.144 \pm 0.006\stat \pm 0.014\syst =  -0.144 \pm 0.015$, 
is in good agreement with the SLD, LEP and ATLAS results.  
It is also compatible with the world average value of the lepton asymmetry parameter $A_{\ell}$~\cite{Rpp:2022}. 
This result is at present the most precise measurement at hadron colliders 
and reaches a similar precision to the SLD experiment.

The measured polarization constrains the effective couplings of \tauleptonsminus to the \PZ boson and determines the effective weak mixing angle 
to be $\sintwoth=0.2319 \pm 0.0019$.  This result has a precision of 0.8\% and is independent of the production process of the \PZ boson.

\begin{acknowledgments}
  We congratulate our colleagues in the CERN accelerator departments for the excellent performance of the LHC and thank the technical and administrative staffs at CERN and at other CMS institutes for their contributions to the success of the CMS effort. In addition, we gratefully acknowledge the computing centers and personnel of the Worldwide LHC Computing Grid and other centers for delivering so effectively the computing infrastructure essential to our analyses. Finally, we acknowledge the enduring support for the construction and operation of the LHC, the CMS detector, and the supporting computing infrastructure provided by the following funding agencies: SC (Armenia), BMBWF and FWF (Austria); FNRS and FWO (Belgium); CNPq, CAPES, FAPERJ, FAPERGS, and FAPESP (Brazil); MES and BNSF (Bulgaria); CERN; CAS, MoST, and NSFC (China); MINCIENCIAS (Colombia); MSES and CSF (Croatia); RIF (Cyprus); SENESCYT (Ecuador); MoER, ERC PUT and ERDF (Estonia); Academy of Finland, MEC, and HIP (Finland); CEA and CNRS/IN2P3 (France); SRNSF (Georgia); BMBF, DFG, and HGF (Germany); GSRI (Greece); NKFIH (Hungary); DAE and DST (India); IPM (Iran); SFI (Ireland); INFN (Italy); MSIP and NRF (Republic of Korea); MES (Latvia); LAS (Lithuania); MOE and UM (Malaysia); BUAP, CINVESTAV, CONACYT, LNS, SEP, and UASLP-FAI (Mexico); MOS (Montenegro); MBIE (New Zealand); PAEC (Pakistan); MES and NSC (Poland); FCT (Portugal); MESTD (Serbia); MCIN/AEI and PCTI (Spain); MOSTR (Sri Lanka); Swiss Funding Agencies (Switzerland); MST (Taipei); MHESI and NSTDA (Thailand); TUBITAK and TENMAK (Turkey); NASU (Ukraine); STFC (United Kingdom); DOE and NSF (USA).
    
  \hyphenation{Rachada-pisek} Individuals have received support from the Marie-Curie program and the European Research Council and Horizon 2020 Grant, contract Nos.\ 675440, 724704, 752730, 758316, 765710, 824093, and COST Action CA16108 (European Union); the Leventis Foundation; the Alfred P.\ Sloan Foundation; the Alexander von Humboldt Foundation; the Science Committee, project no. 22rl-037 (Armenia); the Belgian Federal Science Policy Office; the Fonds pour la Formation \`a la Recherche dans l'Industrie et dans l'Agriculture (FRIA-Belgium); the Agentschap voor Innovatie door Wetenschap en Technologie (IWT-Belgium); the F.R.S.-FNRS and FWO (Belgium) under the ``Excellence of Science \NA EOS" \NA be.h project n.\ 30820817; the Beijing Municipal Science \& Technology Commission, No. Z191100007219010 and Fundamental Research Funds for the Central Universities (China); the Ministry of Education, Youth and Sports (MEYS) of the Czech Republic; the Shota Rustaveli National Science Foundation, grant FR-22-985 (Georgia); the Deutsche Forschungsgemeinschaft (DFG), under Germany's Excellence Strategy \NA EXC 2121 ``Quantum Universe" \NA 390833306, and under project number 400140256 \NA GRK2497; the Hellenic Foundation for Research and Innovation (HFRI), Project Number 2288 (Greece); the Hungarian Academy of Sciences, the New National Excellence Program \NA \'UNKP, the NKFIH research grants K 124845, K 124850, K 128713, K 128786, K 129058, K 131991, K 133046, K 138136, K 143460, K 143477, 2020-2.2.1-ED-2021-00181, and TKP2021-NKTA-64 (Hungary); the Council of Science and Industrial Research, India; the Latvian Council of Science; the Ministry of Education and Science, project no. 2022/WK/14, and the National Science Center, contracts Opus 2021/41/B/ST2/01369 and 2021/43/B/ST2/01552 (Poland); the Funda\c{c}\~ao para a Ci\^encia e a Tecnologia, grant CEECIND/01334/2018 (Portugal); the National Priorities Research Program by Qatar National Research Fund; MCIN/AEI/10.13039/501100011033, ERDF ``a way of making Europe", and the Programa Estatal de Fomento de la Investigaci{\'o}n Cient{\'i}fica y T{\'e}cnica de Excelencia Mar\'{\i}a de Maeztu, grant MDM-2017-0765 and Programa Severo Ochoa del Principado de Asturias (Spain); the Chulalongkorn Academic into Its 2nd Century Project Advancement Project, and the National Science, Research and Innovation Fund via the Program Management Unit for Human Resources \& Institutional Development, Research and Innovation, grant B05F650021 (Thailand); the Kavli Foundation; the Nvidia Corporation; the SuperMicro Corporation; the Welch Foundation, contract C-1845; and the Weston Havens Foundation (USA).  
\end{acknowledgments}

\bibliography{auto_generated}
\cleardoublepage \appendix\section{The CMS Collaboration \label{app:collab}}\begin{sloppypar}\hyphenpenalty=5000\widowpenalty=500\clubpenalty=5000
\cmsinstitute{Yerevan Physics Institute, Yerevan, Armenia}
{\tolerance=6000
A.~Hayrapetyan, A.~Tumasyan\cmsAuthorMark{1}\cmsorcid{0009-0000-0684-6742}
\par}
\cmsinstitute{Institut f\"{u}r Hochenergiephysik, Vienna, Austria}
{\tolerance=6000
W.~Adam\cmsorcid{0000-0001-9099-4341}, J.W.~Andrejkovic, T.~Bergauer\cmsorcid{0000-0002-5786-0293}, S.~Chatterjee\cmsorcid{0000-0003-2660-0349}, K.~Damanakis\cmsorcid{0000-0001-5389-2872}, M.~Dragicevic\cmsorcid{0000-0003-1967-6783}, A.~Escalante~Del~Valle\cmsorcid{0000-0002-9702-6359}, P.S.~Hussain\cmsorcid{0000-0002-4825-5278}, M.~Jeitler\cmsAuthorMark{2}\cmsorcid{0000-0002-5141-9560}, N.~Krammer\cmsorcid{0000-0002-0548-0985}, D.~Liko\cmsorcid{0000-0002-3380-473X}, I.~Mikulec\cmsorcid{0000-0003-0385-2746}, J.~Schieck\cmsAuthorMark{2}\cmsorcid{0000-0002-1058-8093}, R.~Sch\"{o}fbeck\cmsorcid{0000-0002-2332-8784}, D.~Schwarz\cmsorcid{0000-0002-3821-7331}, M.~Sonawane\cmsorcid{0000-0003-0510-7010}, S.~Templ\cmsorcid{0000-0003-3137-5692}, W.~Waltenberger\cmsorcid{0000-0002-6215-7228}, C.-E.~Wulz\cmsAuthorMark{2}\cmsorcid{0000-0001-9226-5812}
% ...
\par}
\vskip\cmsinstskip
\dag:~Deceased\\
$^{1}$Also at Yerevan State University, Yerevan, Armenia\\
$^{2}$Also at TU Wien, Vienna, Austria\\
% ...
\end{sloppypar}
\end{document}